\documentclass[twocolumn,letterpaper,10pt]{IEEEtran}
\usepackage[margin=0.75in]{geometry}
\usepackage{graphicx}
\usepackage{subfigure}
\usepackage{amsmath,amssymb}\usepackage{amsthm}
\usepackage[mathscr]{eucal}
\usepackage{graphics,graphicx,multicol}
\usepackage{color}
\usepackage{amsfonts}
\usepackage{epsfig}
\usepackage{cite,xspace,syntonly,algorithm,algorithmic}
\usepackage{bm}

\newcommand{\Reals}     {{{\mathrm I\!R}}}  

\newcommand{\define}    {\stackrel{\scriptscriptstyle\triangle}{=}}  
\newcommand{\diag}    {{\mathrm{diag}}}  

\newcommand{\Zrb}     {{\uwti 0}}      
\newcommand{\Oneb}    {{\uwti 1}}      

\newcommand{\uwti}[1]{{\mathbf #1}}
  \newcommand{\Ab}{{\uwti A}}
  
  \newcommand{\Cb}{{\uwti C}}
  \newcommand{\Db}{{\uwti D}}
  
  \newcommand{\Fb}{{\uwti F}}
  \newcommand{\Gb}{{\uwti G}}
  \newcommand{\Hb}{{\uwti H}}
  \newcommand{\Ib}{{\uwti I}}

  \newcommand{\Lb}{{\uwti L}}

\renewcommand{\sb}{{\uwti s}} 
  \newcommand{\Tb}{{\uwti T}}
  
  \newcommand{\Vb}{{\uwti V}}
\newcommand{\wb}{{\uwti w}}  \newcommand{\Wb}{{\uwti W}}
\newcommand{\xb}{{\uwti x}}  
\newcommand{\yb}{{\uwti y}}  
\newcommand{\zb}{{\uwti z}}  

          \newcommand{\Gammab}   {{\bm \Gamma}}

\newcommand{\zetab}       {{\bm \zeta}}           
\newcommand{\etab}        {{\bm \eta}}

          \newcommand{\Sigmab}   {\uwti{\mathnormal\Sigma}}

\newcommand{\psib}        {{\bm \psi}}            \newcommand{\Psib}     {{\bm \Psi}}

\newcommand{\Ac} {{\mathcal A}}         
\newcommand{\Bc} {{\mathcal B}}         
\newcommand{\Cc} {{\mathcal C}}         
\newcommand{\Dc} {{\mathcal D}}         
\newcommand{\Ec} {{\mathcal E}}         
\newcommand{\Fc} {{\mathcal F}}

\newcommand{\Qc} {{\mathcal Q}}         
         
\newcommand{\Sc} {{\mathcal S}}         
\newcommand{\Tc} {{\mathcal T}}         
\newcommand{\Uc} {{\mathcal U}}         
         
         \newcommand{\Wk} {{\bm {\mathcal W}}}

\newcommand{\Pulk} {{\underline{{\bm {\mathcal P}}}}}

\newcommand{\Iulk} {{\underline{{\bm {\mathcal I}}}}}

\newcommand{\Gulc} {{\underline{\mathcal G}}}

              \newcommand{\eul}  {{\underline e}}

\newcommand{\Iul}  {{\underline I}}

\newcommand{\Omegaul}  {{\underline \Omega}}
\newcommand{\ba}{\begin{array}}
\newcommand{\ea}{\end{array}}

 \newtheorem{definition}{Definition}
\newtheorem{coro}{Corollary}
\newtheorem{remark}{Remark}

\newtheorem{lemma}{Lemma}
\newtheorem{prop}{Proposition}

\begin{document}
\pagestyle{empty}
\title{ \vspace{.5cm} Optimizing Beams and Bits: A Novel Approach for Massive MIMO Base-Station Design}
\author{\IEEEauthorblockN{Narayan Prasad\IEEEauthorrefmark{1},
Xiao-Feng Qi\IEEEauthorrefmark{1} and Alan Gatherer\IEEEauthorrefmark{2}}\\
\IEEEauthorblockA{\IEEEauthorrefmark{1}Futurewei Technologies, Radio Algorithms Research Group, NJ Research Center, Bridgewater, NJ USA\\
\IEEEauthorrefmark{2}Futurewei Technologies,  TX USA}\\
e-mail:   \{narayan.prasad1, xiao.feng.qi, alan.gatherer\}@huawei.com
\vspace{-2ex}}
\maketitle \vspace{-.1cm}
\thispagestyle{empty}
 

%
%
%
%
%
%
\begin{abstract}
We consider the problem of jointly optimizing ADC bit resolution and 
   analog beamforming over a frequency-selective  massive MIMO  uplink. We build upon a popular model to incorporate the impact of low bit resolution ADCs, that hitherto has mostly been employed over flat-fading  systems. We adopt   weighted sum rate (WSR) as our objective and show that WSR maximization under finite buffer limits and important practical constraints on choices of beams and ADC bit resolutions can  equivalently be posed as constrained  submodular set function maximization. This enables us to design a constant-factor approximation algorithm. Upon incorporating further enhancements we obtain an efficient algorithm that significantly outperforms state-of-the-art ones. 
 
\end{abstract}

\section{Introduction}
%
In this paper
we consider a critical issue impacting next generation (5G and beyond) cellular deployments. It is well recognized that massive MIMO is a key 5G technology that promises very substantial throughput improvements, at-least in the presence of accurate channel state information \cite{larsson:2014,hoydis}. However, cost considerations at both the deployment stage (capex) as well operational stage (opex) have raised several concerns on large scale adoption of this technology. Indeed, the number of RF chains must be limited to keep capex viable, and power consumption needs to be curtailed both from operational expenditure and environmental impact points of view. 
Recent research has increasingly focused on hybrid architectures that can potentially capture a substantial portion of available gains using much fewer RF chains. Simultaneously, adaptive resolution analog-to-digital-converters (ADCs) have also received wide attention as a means to significantly cut down power consumption \cite{Mo17,RothNoss17,choi17}.

Our focus here is to establish a sound theoretical framework for optimally exploiting both hybrid architecture and adaptive ADC.  The setting we choose is a practical wideband frequency-selective uplink incorporating multi-path in the propagation and OFDMA as the multiple access scheme. The objective we seek to maximize via joint optimization is the (queue-constrained) weighted sum-rate (WSR) metric. WSR metric is the paramount objective  in resource allocation at fine time scales, since by adapting the weights appropriately one can enforce any desired policy over longer time-scales.  
The model we rely on to incorporate impact of quantization is based on a simplified approach that comprises of scaling the input and adding a quantization noise term  \cite{Orhan17}. This approach (referred to as AQNM) has been effectively exploited previously in \cite{choi17,choiAS18,Abbas:mmWADC}, mostly over flat-fading systems, with   a notable recent exception being \cite{Roth:WBHybrid}, which exploits AQNM over a wideband uplink.  By leveraging AQNM we systematically obtain a model for the wideband uplink by highlighting all key steps and assumptions. The resulting model explicitly includes quantization effects and is tractable in that it facilitates sophisticated optimization techniques that seek to maximize  WSR  metric. 
 To the best of our knowledge, this paper is the first to consider quantization-aware queue-constrained WSR optimization  over   frequency selective systems. Notable recent works have focused mainly on a flat-fading model  and other objectives   (such as mean squared error in \cite{choi17}) or sum rate \cite{choiAS18,RothNoss17,Abbas:mmWADC}, with \cite{choiAS18} considering receive antenna subset selection which is a special case of beam group optimization  (with fixed ADC resolution). We note that the recent contribution in \cite{Roth:WBHybrid} does consider a frequency selective uplink with two levels for  ADC bit resolutions and  analog beam group selection. However,   joint optimization is not rigorously pursued there, with the   criterion used for beam group selection being based on received power (without considering  impact of subsequent quantization).  Also noteworthy are \cite{studerWB} and \cite{MollenWB}  both of which consider low-bit resolution ADCs over a frequency selective uplink. Specifically, \cite{studerWB} focuses on MAP and other more tractable data detection schemes,  whereas \cite{MollenWB} derives  achievable rate expressions for 1 bit ADC under different asymptotic regimes.  

Our main contribution in this paper is to cast the constrained joint maximization of WSR  {\em  as a discrete submodular set function maximization problem.}
 Using this route of discrete optimization confers several advantages since the original problem at hand is inherently a  discrete optimization over analog codebook subsets and ADC bit resolutions. Indeed, we now no longer have to relax the bits to be continuous variables and we can use any arbitrary look-up-table to obtain  effective quantization gain as a function of ADC bit resolution. A similar comment applies with respect to the energy cost of operating an RF chain with an ADC at any chosen  bit resolution.  In this context, we note that proper modeling of quantization gains and energy costs is essential to obtain true gains.   
 {\em Our work   recognizes that submodularity can be  exploited in the joint optimization of analog beam group and ADC bit resolutions even after explicitly modeling quantization impact.} 
This allows us to derive a constant-factor approximation algorithm\footnote{Such an algorithm guarantees that the WSR it yields will be at-least a constant-fraction of the optimal WSR for every input instance.}.  We then recast our problem using submodular cost constraints and obtain a low complexity enhanced algorithm that leverages the special structure present in our re-formulated optimization problem. {\em Consequently, we are able to demonstrate significant performance gains  even with a  reduced complexity compared to state-of-the-art schemes \cite{choiAS18}.} Indeed, we  show that our enhanced  algorithm yields  upto   $50\%$  WSR gains over other schemes in a regime with tight power (energy) budgets.  
At the same time, our algorithm can match or exceed the near-optimal throughput performance of other schemes albeit with $40-{\rm to}-50 \%$ reduction in consumed energy. 
  
 Over the past decade results establishing submodularity for  a variety of problems are increasingly available. These include sensor placement,   single-user  scheduling (that schedules users on orthogonal time-frequency resources) with   fixed transmit powers    \cite{andrews07} as well as optimized powers   \cite{thekum:wf}. Submodularity has also been shown to hold in  formulations considering the user and base-station association problem  \cite{singh:af,Aopsounis}, caching \cite{Golrezaei} and to some extent even multi-user MIMO scheduling (that schedules multiple users on same time-frequency resource) \cite{prasad:dpc}. The main motivation for these works is the availablitly of increasingly effective approximation algorithms for constrained   submodular set function maximation \cite{calinescu09,gamzu2012,Aclark}. 
Our work here adds to this growing body of knowledge by establishing submodularity for a problem where the impact of imperfect (finite resolution) quantization is explicitly modeled, and also by deriving an effective approximation algorithm that considers submodular constraints.
%

\section{System Model}
\label{sect:formulation}
\begin{figure}[t]
 \begin{minipage}[t]{\linewidth}
    \centering
    \includegraphics[width=1\linewidth]{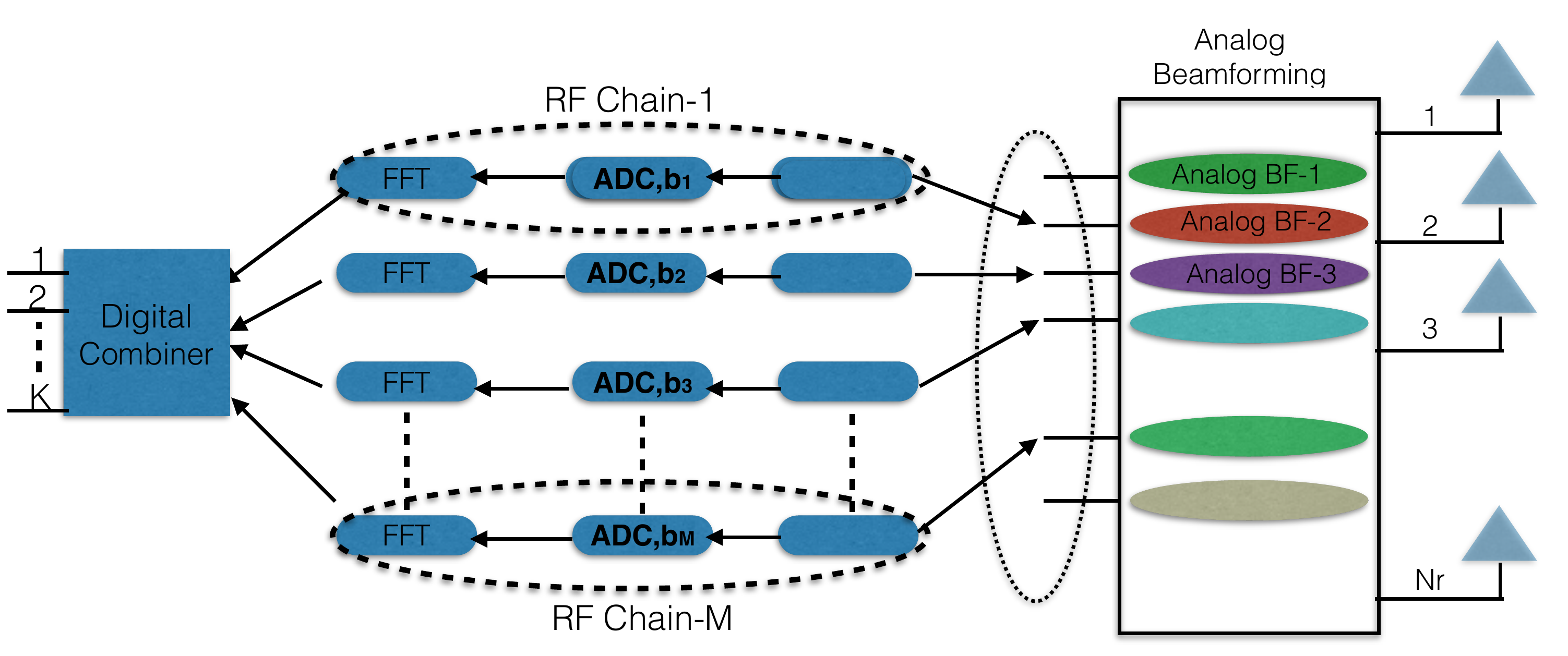}
    \vspace{-1cm}\caption{System Schematic}\label{fig:SysSch}
\end{minipage}
\end{figure}

We focus on a single-cell uplink that comprises of a base station (BS)  which is equipped with a large array of $N_r$ ($N_r>>1$) receive antennas. Due to cost restrictions the BS has a fewer number,  $M: M\leq N_r$, of RF chains. The BS communicates with $K$ users, with each user being equipped with a single transmit antenna. Suppose   the uplink access scheme to be OFDMA and let $N$ denote the number of subcarriers.
Further assume that the transmit powers used by all users on all subcarriers are given as  inputs. In addition, the queue size (amortized to per symbol) and the weight of each user $k$, denoted by $Q_k$ and $w_k$, respectively, are also specified.
Our objective here is to determine a rate assignment that maximizes queue-constrained  WSR $\sum_{k=1}^Kw_kR_k$ among all achievable rate assignments, where $R_k\;\forall\;k$ denotes the rate assigned to user $k$ in bits per OFDM symbol (satisfying $R_k\leq Q_k$). The set of achievable rate assignments $\{[R_1,\cdots,R_K]\}$ depends on certain BS receiver attributes that are illustrated by the system schematic in Fig. \ref{fig:SysSch}. 
The diagram in Fig. \ref{fig:SysSch} assumes that an analog beamforming  codebook is employed at the BS receiver. Using this codebook the BS can activate any subset of up-to $M$ analog beamformers and connect the output of each selected beamformer to a (unique) RF chain.\footnote{Note that the switching setup is shown in Fig. \ref{fig:SysSch}  for convenience. Each RF chain can instead also employ a beam formed via a dedicated bank of finite resolution phase shifters. } Each RF chain also houses an ADC whose bit resolution can be configured. The set of achievable rate assignments thus depends on the  subset of chosen analog beamformers as well as the bit resolutions configured for the ADCs on  RF chains those beam outputs are connected to.

Let us proceed to formally specify a system model. Consider any  analog beamforming codebook comprising of a set of orthonormal analog beams. Suppose that $M$ analog beams are chosen which activates all $M$ available RF chains. For our purposes here the  mapping between activated  RF chains and   outputs post-analog beamforming  is not important, as long as it is one-to-one.  
Then, let    $\Wb$ denote an $M\times N_r$ matrix   whose rows comprise of  $M$ selected (mutually orthogonal and unit-norm) analog beamforming vectors, so that $\Wb\Wb^\dag=\Ib$.  
Next,  model the received vector  at the input of the bank of  ADCs via the standard baseband multi-user MIMO-OFDMA \cite{tseVbook} time-domain input output relation  as
\begin{eqnarray}\label{eq:chmod}
\yb = \Hb \xb +\etab,
\end{eqnarray}
where $\yb=[\yb_1^T,\cdots,\yb_N^T]^T$ denotes the $NM\times 1$ vector of observations  received over $N$ chip durations (equivalently over the OFDM symbol duration). 
Notice here that the observations in $\yb$ are post analog beamforming and  after removal of the cyclic prefix. 
$\etab$ denotes the additive    circularly symmetric complex Gaussian noise vector with $E[\etab\etab^\dag]=\Ib$. 
The vector $\xb=[\xb_1^T,\cdots,\xb_N^T]^T$ is the $NK\times 1$ vector of time-domain transmissions from the $K$ users. Recall that each user obtains its time-domain   signal by applying an inverse DFT matrix to an $N$ length frequency domain symbol vector. Thus, we can express $\xb$ as
 $\xb = (\Fb^\dag\otimes \Ib_K) \sb$, where $\Fb$ is an $N\times N$  DFT matrix so that $\Fb^\dag$ is its inverse.  We parse the   circularly symmetric complex normal vector $\sb$ as $\sb = [\sb_1^T,\cdots,\sb_N^T]^T$ where $\sb_n  = [s_{n,1},\cdots, s_{n,K}]^T,\;1\leq n\leq N$ denotes the  vector of symbols transmitted by the  $K$ users on the $n^{th}$ subcarrier.  Then, we let $\Db_n=E[\sb_n\sb_n^\dag], 1\leq n\leq N$ denote the given (power loading) diagonal covariance matrix on the $n^{th}$ subcarrier. We note here that we allow for $\Db_n$ to be any given diagonal positive semi-definite matrix. Further, we form the matrix $\Db=E[\sb\sb^\dag]$ such that $\Db = {\rm BlkDiag}\{\Db_1,\cdots,\Db_N\}$ is a  diagonal  matrix whose $n^{th}$ diagonal block is $\Db_n$. 
Finally, the matrix $\Hb$ in (\ref{eq:chmod}) representing the effective channel post analog beamforming is an $NM \times NK$  block circulant matrix.  To specify $\Hb$  we expand it in terms of its constituent $N$ blocks as 
\begin{eqnarray}
\Hb = 
\left[
\begin{array}{cccc}
\Hb_{(1)} & \Hb_{(2)} &\cdots & \Hb_{(N)} \\
\Hb_{(N)} & \Hb_{(1)} &\cdots & \Hb_{(N-1)}\\
\vdots & \ddots &\cdots & \vdots \\
\Hb_{(2)} & \Hb_{(3)} &\cdots & \Hb_{(1)}
\end{array}
\right].
\end{eqnarray}
Consequently, it suffices to specify the first block row of $\Hb$, which in turn is given by  
\begin{eqnarray}
\nonumber [\Hb_{(1)},\Hb_{(2)},\cdots,\Hb_{(N-L+1)},\Hb_{(N-L+2)},\cdots,\Hb_{(N)}] =\\
 \Wb[\Hb_0,\Zrb,\cdots,\Zrb,\Hb_{L-1},\cdots,\Hb_1], 
\end{eqnarray}
 where we note that each one of  the matrices $\Hb_{(k)},\;2\leq k\leq N-L+1$ has  all of its entries to be zero.
Further the matrix  $\Hb_i,\;0\leq i\leq L-1$ 
 denotes the $N_r\times K$ matrix modeling the $(i+1)^{th}$ tap (or path) and $L$ is the number of paths.  We assume that accurate estimates of these per-tap matrices are available at the BS.  \footnote{With adaptive ADCs we can set each ADC resolution to be highest possible during channel estimation phase which somewhat justifies assuming availability of accuarate channel estimates at the BS. } Notice that without loss of generality we have assumed an identical number of paths for all users. This is because we can always choose $L$ (and cyclic prefic length) based on the user corresponding to the largest delay spread and then use zero-padding.
 \subsection{Modeling Quantization}\label{sec:quantmod}
We are now ready to consider quantization performed by the bank of ADCs. We assume that each ADC independently quantizes only the input received by it (scalar quantization). Accordingly, let $\yb^{(q)}$ denote the vector after element-wise quantization of $\yb$  in (\ref{eq:chmod}). Note further that each ADC is in-fact a pair quantizing the real and imaginary parts separately (both using the assigned bit resolution). 
We will adopt a particular   additive quantization noise model (AQNM) which bestows tractability while being relevant \cite{Orhan17}.  This particular AQNM model has   been effectively adopted recently in \cite{Mo17,Abbas:mmWADC,choi17}  and its accuracy improves at low to moderate SNRs \cite{RothNoss17}. 
Noting that for a given channel realization and analog beamforming matrix, $\yb$ is a zero mean complex normal vector, the key entities we need to determine in order to employ the said AQNM model are variances $E[|y_j|^2],\;1\leq j\leq NM$, where $y_j$ is the $j^{th}$ entry of $\yb$.  
We have the following result which follows from some careful algebra and states that for each analog beamformer output these variances are identical across time.
\begin{lemma}\label{lem1}
Let $\Cb=E[\yb\yb^{\dag}]$ denote the covariance matrix of  the input to the quantizer bank. Then, $\Cb$ is a block circulant matrix with $M\times M$ constituent blocks.  The $MN$ diagonal elements of $\Cb$ can be expressed as 
\begin{eqnarray}\label{eq:diagC}
 \Oneb\otimes \psib,\;\rm{ with}\;
\psib = [\psi_1,\cdots,\psi_{M}]^T, 
\end{eqnarray}
where $\Oneb$ denotes the $N\times 1$ vector of all ones and  $\psi_m,\;1\leq m\leq M$ denotes the identical variance of all the outputs corresponding to the $m^{th}$ analog beamforming vector across time.   Further, the entries of 
 $\psib$ are  the diagonal elements of the $M\times M$ matrix:\\
$\Ib + \Wb[\Hb_0,\Zrb,\cdots,\Zrb,\Hb_{L-1},\cdots,\Hb_1](\Fb^\dag\otimes\Ib_K)\Db(\Fb\otimes\Ib_K)[\Hb_0,\Zrb,\cdots,\Zrb,\Hb_{L-1},\cdots,\Hb_1]^\dag\Wb^\dag$.  \\
Thus, for each $m: 1\leq m\leq M$, 
 $\psi_m$ is invariant to  choice of  all analog beamforming vectors other than the $m^{th}$ one. 
\end{lemma}
 
Now we are ready to model the vector $\yb^{(q)}$ obtained post quantization. Suppose that the ADC for the $m^{th}$ analog beam vector output is assigned a resolution of $b_m$ bits.  For the given resolution $b_m$, we define $M$ positive quantization scalars $\alpha_m\in [0,1],\;1\leq m\leq M$. A popular choice (which corresponds to scalar non-uniform mmse quantizer for Gaussian inputs) is to set $\alpha_m = 1-a2^{-2b_m}$ whenever $b_m>5$ (for some positive constant $a$). On the other hand,  $\alpha_m$ is read from  a look-up-table for $b_m=1,\cdots,5$. For our purposes, we can employ any arbitrarily specified look-up-table  to read $\alpha_m$ as a function of $b_m$ so long as
 $\alpha_m$ is  increasing in $b_m$.   Building upon the simplified  AQNM \cite{Orhan17}, we assume $\yb^{(q)}$ can be expanded as 
\begin{eqnarray*}
\nonumber \yb^{(q)}  
 = \Ab \Hb\xb +  \breve{\etab}^{(q)}
\end{eqnarray*}
 where $\Ab =\Ib_N\otimes \diag\{\alpha_1,\cdots,\alpha_M\}$ and $\breve{\etab}^{(q)}$ is the total noise (including additional quantization noise) that is zero-mean and uncorrelated to   $\xb$. 
 Next, following standard OFDM processing at the BS,   $\yb^{(q)}$ is transformed by using  a DFT matrix on the outputs corresponding to each analog beam separately, i.e., we obtain
\begin{eqnarray}\label{eq:expmod}
\nonumber \zb   &=& (\Fb\otimes\Ib_M)\yb^{(q)} =  (\Fb\otimes\Ib_M)\Ab \Hb\xb +    (\Fb\otimes\Ib_M)\breve{\etab}^{(q)}\\
 \nonumber &=&  (\Fb\otimes\Ib_M)\Ab \Hb(\Fb^\dag\otimes \Ib_K) \sb +    (\Fb\otimes\Ib_M)\breve{\etab}^{(q)}\\
 &=& \Ab\Gb\sb +     \tilde{\etab}^{(q)}
\end{eqnarray}
where for the last equality we have defined $ \tilde{\etab}^{(q)}\define (\Fb\otimes\Ib_M)\breve{\etab}^{(q)}$ and used the fact that
 $(\Fb\otimes\Ib_M)\Ab \Hb(\Fb^\dag\otimes \Ib_K)=\Ab\Gb$.  $\Gb={\rm BlkDiag}\{\Wb\Gb_1,\cdots,\Wb\Gb_N\}$ is a block diagonal  matrix whose $n^{th}$ diagonal block is given
 by $\Wb\Gb_n = \Wb\sum_{\ell=0}^{L-1}\Hb _\ell \exp(-j2\pi (n-1)\ell/N),\;1\leq n\leq N$. 
Analogous to typical modeling (cf. \cite{Roth:WBHybrid}) we also suppose the transformed total noise in the frequency domain,
 $\tilde{\etab}^{(q)}$, to be a circularly symmetric complex normal vector that is  independent of $\sb$. 

 A key factor that will determine the extent of tractability of  (\ref{eq:expmod}) is   the form of the covariance 
  of $\tilde{\etab}^{(q)}$. 
  If we further follow the simplified AQNM assumptions, we will first obtain that
$E[\breve{\etab}^{(q)}(\breve{\etab}^{(q)})^\dag]=\Ab^2 + \Ab(\Ib - \Ab)\Psib$ where $\Psib$ is an $MN\times MN$  diagonal matrix whose diagonal  elements   are identical to the variances of the corresponding quantizer inputs, i.e., identical to the respective diagonal elements of $\Cb=E[\yb\yb^\dag]$ . Then, invoking Lemma \ref{lem1} and in particular the special structure of the diagonal elements of $\Cb$ in (\ref{eq:diagC}), yields that $\Psib=\Ib_N\otimes\diag\{\psib\}$. 
 This in turn results in
\begin{eqnarray}\label{eq:covmod}
E[ \tilde{\etab}^{(q)} ( \tilde{\etab}^{(q)})^{\dag}] = \Ab^2 + \Ab(\Ib-\Ab)\Psib = \Ib_N\otimes\Gammab
\end{eqnarray}
 wherein $\Gammab=\diag\{\gamma_1,\cdots,\gamma_M\}$ is an $M\times M$ diagonal matrix whose $m^{th}$ diagonal element is given by $\gamma_m = \alpha_m^2+\alpha_m(1-\alpha_m)\psi_m,\;1\leq m\leq M$.   
Now, expanding $\zb$ in (\ref{eq:expmod}) in terms of its per-subcarrier components, we see that tractability holds. This is because the covariance derived in (\ref{eq:covmod}) implies that noise across different subcarriers is uncorrelated. Then, upon whitening the total noise on each subcarrier we obtain our desired model 
\begin{eqnarray}\label{eq:chmodsp}
 \tilde{\zb}_n   = \Tb^{1/2}\Wb\Gb_n\sb_n +     \zetab_n,\;1\leq n\leq N,
\end{eqnarray}
where $E[\zetab_n\zetab_n^\dag]=\Ib$, with $E[\zetab_n\zetab_m^\dag]=\Zrb\;\forall n\neq m$, and the diagonal matrix $\Tb=\diag\{\alpha_1^2/\gamma_1,\cdots,\alpha_M^2/\gamma_M \}$  is invariant across all subcarriers. 
 Notice  that (\ref{eq:chmodsp}) is a wideband model incorporating multi-path in propagation and quantization at the receiver. Specializing (\ref{eq:chmodsp}) to the single-path flat fading case ($L=1$),  we recover the narrowband model of \cite{choi17}.
 
We remark here that a  more  general (finer) modeling is one under which   total noise covariance  $E[\breve{\etab}^{(q)}(\breve{\etab}^{(q)})^\dag]$ is   approximated by
  any positive definite $MN\times MN$ block circulant matrix whose constituent $M\times M$ blocks are all diagonal. Clearly 
 the choice $E[\breve{\etab}^{(q)}(\breve{\etab}^{(q)})^\dag]=\Ab^2 + \Ab(\Ib - \Ab)\Psib$ is a special case under this more general framework wherein we further set all off-diagonal blocks to be zero. 
Indeed, the more general framework also  maintains tractability and yields a per-subcarrier model   
 \begin{eqnarray}\label{eq:chmodsp2}
 \tilde{\zb}_n   = \Tb_n^{1/2}\Wb\Gb_n\sb_n +     \zetab_n,\;1\leq n\leq N,
\end{eqnarray}
where each $\Tb_n=\diag\{T_{n1},\cdots,T_{nM}\}\forall\;n$ and $E[\zetab_n\zetab_n^\dag]=\Ib\;\forall n$ with $E[\zetab_n\zetab_m^\dag]=\Zrb\;\forall n\neq m$ as before. In the sequel, for notational simplicity we consider the  model in (\ref{eq:chmodsp}) but note that all our results immediately extend to the one in (\ref{eq:chmodsp2}), so long as two natural conditions are satisfied by the model in (\ref{eq:chmodsp2}). These conditions, which are both met by the special case in (\ref{eq:chmodsp}), are: 
(i)  for each $m$, $\{T_{nm}\}\;\forall n$, must not depend on the choice of bit resolutions or analog beamformers  made for chains other than the $m^{th}$ one. (ii) For each $m$,   $\{T_{nm}\}\;\forall n$ are all positive and monotonically increasing in the $m^{th}$ ADC bit resolution,  $b_m$.

\section{Joint Analog Beams and  Bit Resolutions Optimization via Submodular optimization}
In this section we will jointly optimize the choice of  analog beams and bit resolutions of their corresponding ADCs. Towards this goal, 
we suppose that the set of all available $N_r$  length analog beam vectors, denoted by $\Wk$, is finite\footnote{ This is a practical case where BS employs a finite codebook of beams.} and comprises of mutually orthogonal  beam vectors so that $|\Wk|\leq N_r$. Similarly, the set of all possible (strictly positive) bit resolutions that we are allowed to assign to quantize the output of any selected beam is also assumed to be finite and is denoted by  $\Bc$.
Recall that  $Q_k$  denotes the given number of bits in the queue of user $k\in\Uc$  where $\Uc=\{1,\cdots,K\}$ is the user pool.  
Furthermore, in order to target the best possible performance that can be obtained using analog receive beamforming and ADCs with adaptive bit resolution, for any choice of analog receive beam vectors and ADC bit resolutions, we assume that the subsequent decoding at the BS is optimal.  Thus, all beam outputs post-quantization are used to {\em jointly}  decode all user signals. 
We accordingly define a ground set   comprising of all possible tuples or pairs, where each such tuple denotes a selection of an analog beam and a bit resolution for its associated ADC. In  particular, we define the ground set as $\Omegaul = \{(\wb,b):\wb\in\Wk\;\&\;b\in\Bc \}$ so that  its cardinality equals $|\Omegaul|=|\Wk| |\Bc|$. 
Then for any  choice of subset $\Gulc \subseteq\Omegaul$ of tuples, we have a set of analog receive beams and bit resolutions specified in those tuples. 
Note that when the beams across all tuples of $\Gulc$ are distinct, they must be mutually orthogonal (since any two beams in $\Wk$ are mutually orthogonal).  To enforce that a feasible choice of $\Gulc$   includes each beam in at-most one of its tuples, we can define 
  $\Iulk$ to denote a family of subsets of $\Omegaul$ such that: each 
 member in $\Iulk$ contains only distinct beams across its constituent tuples and any subset of $\Omegaul$ in which the constituent tuples have distinct beams is a member of $\Iulk$. The family $\Iulk$ defined this way can be seen to be a  matroid (cf. definitions given in the appendix).  Then, for any $\Gulc\in\Iulk$, 
using the   beams and bit resolutions in $\Gulc$ we can form the matrix $\Wb$ and determine the matrix $\Tb$ in (\ref{eq:chmodsp}), where we note that   $M$ must now be replaced by $|\Gulc|$. To explicitly indicate the dependence on $\Gulc$,  we will denote the corresponding matrices by $\Wb_\Gulc$ and $\Tb_\Gulc$, respectively, where $\Wb_\Gulc$ is a $|\Gulc|\times N_r$ matrix while $\Tb_\Gulc$ is a $|\Gulc|\times|\Gulc|$ diagonal matrix. Indeed, for each tuple $(\wb,b)\in\Gulc$, the beam $\wb$ is present as a row of $\Wb_\Gulc$. The corresponding diagonal element of $\Tb_\Gulc$ can be computed as $\frac{\alpha^2}{\alpha^2+\alpha(1-\alpha)\psi}$, where  the quantization scalar $\alpha$ is obtained using the given look-up-table and the bit resolution $b$ specified for beam $\wb$ in tuple $(\wb,b)$ of $\Gulc$. The scalar $\psi$ is the variance of the time-domain outputs that would be seen along beam $\wb$, which is  given by (cf. Lemma \ref{lem1}), $1 + \wb[\Hb_0,\Zrb,\cdots,\Zrb,\Hb_{L-1},\cdots,\Hb_1](\Fb^\dag\otimes\Ib_K)\Db(\Fb\otimes\Ib_K)[\Hb_0,\Zrb,\cdots,\Zrb,\Hb_{L-1},\cdots,\Hb_1]^\dag\wb^\dag$. Using
   $\Wb_\Gulc$ and $\Tb_\Gulc$   we 
  write the model in (\ref{eq:chmodsp}) as
\begin{eqnarray}\label{eq:chmodspN}
 \tilde{\zb}_{\Gulc,n}   = \Tb^{1/2}_\Gulc \Wb_\Gulc\Gb_n\sb_n +     \zetab_{\Gulc,n},\;1\leq n\leq N. 
\end{eqnarray}
Note here that  $\Wb_\Gulc$ and $\Tb_\Gulc$ are invariant across subcarriers and since $\Wb_\Gulc\Wb_\Gulc^\dag=\Ib$,     
our normalization  ensures that 
 $E[\zetab_{\Gulc,n}\zetab_{\Gulc,n}^\dag]=\Ib,\forall\;n$.  
 Let us now proceed to determine the optimal weighted sum rate that can be achieved for any choice of $\Gulc\in\Iulk$.    
 Without loss of generality, let us  suppose that the user weights are ordered as $w_1\geq w_2\geq \cdots\geq w_K$. Considering the model in (\ref{eq:chmodspN}) define the matrices
 $\Lb_{\Gulc,n} \define \Tb_\Gulc^{1/2}\Wb_\Gulc\Gb_n\Db_n^{1/2},\;\forall\;\Gulc\in\Iulk\;\&\;n=1,\cdots,N$. 
For each such matrix, we also adopt the convention that
 $\Lb_{\Gulc,n}^{(\Ac)},\forall\;\Ac\subseteq\Uc=\{1,\cdots,K\}$ denotes the submatrix of $\Lb_{\Gulc,n}$ formed by its  columns with indices in $\Ac$.  
Next, we define several set functions, all of them  over  all subsets of $\Uc$ and   one set function for each group $\Gulc\in\Iulk$, as
\begin{eqnarray}\label{eq:setf}
f^{(\Ac)}_\Gulc = \sum_{n=1}^N\log\left|\Ib + \Lb_{\Gulc,n}^{(\Ac)}(\Lb_{\Gulc,n}^{(\Ac)})^\dag\right|,\;\;\forall\;\Ac\subseteq\{1,\cdots,K\}.
\end{eqnarray}
Note that the model in (\ref{eq:chmodspN}) (under our assumption on noise distribution) represents a vector Gaussian multiple access channel. Thus, for any feasible choice of beams and bit resolutions   $\Gulc\in\Iulk$, $f^{(\Ac)}_\Gulc $ can be recognized to be the maximal sum rate that can be achieved for users in $\Ac$ (in the absence of queue constraints) when the messages of
 other users $\Uc\setminus\Ac$ are known and expurgated \cite{tsepoly}. Recall that in our setting the transmit powers of users on each subcarrier are fixed inputs which  cannot be changed and joint power and rate control is left for future work. We note   that transmit power optimization   is further complicated by the fact that changing user transmit powers even while keeping the choice of beams and bit resolutions fixed,
 can alter the variance of the input  at each ADC and thereby the total noise covariance post-quantization.
 Indeed,   even  switching off some users (binary power control) can reduce the variance of the input  at each ADC and potentially further improve the rates that can be achieved for other users by boosting the effective channels seen by the BS from those users (post-quantization and noise-whitening).  
  Then, since we are interested in the WSR under queue constraints,  we need to define the set of all 
 achievable rate vectors (or assignments) under the given fixed transmit powers.  
 Let $R_k$ denote the rate assigned to user $k\in\Uc$ and define $R_\Ac = \sum_{k\in\Ac}R_k\;\forall\;\Ac\subseteq\Uc$. Then, for any  $\Gulc\in\Iulk$,     the set of all 
 achievable rate vectors is given by 
\begin{eqnarray}
 \Pulk_\Gulc = \left\{ [R_1,\cdots, R_K]\in\Reals_+^K:  \;\; R_\Ac\leq f^{(\Ac)}_\Gulc \;\forall\;\Ac\subseteq\Uc\right\}
\end{eqnarray}
The rate region $\Pulk_\Gulc$ is known to be a polymatroid \cite{tsepoly}. 
To impose the condition that $R_k\leq Q_k\;\forall k$ we only consider rate vectors in $\Pulk_\Gulc$ satisfying these queue constraints. The region formed by all such rate vectors, denoted by $\Pulk'_\Gulc$, can  be shown to be another polymatroid \cite{fujiPoly}.  
Then, we can invoke a fundamental result on polymatroids to deduce  that  {\em a rate vector which   maximizes the weighted sum rate among all vectors in $\Pulk'_\Gulc$, i.e., $\arg\max_{[R_1,\cdots,R_K]\in \Pulk'_\Gulc}\{\sum_{k=1}^Kw_kR_k\}$, is the one corresponding to its corner point determined solely by the assigned user weights \cite{edpoly,fujiPoly}}. This holds true for all choices of the subset $\Gulc$. Therefore, without loss of optimality, we can  associate the weighted sum rate achieved by that corner point as the metric value  for each choice of $\Gulc$. To formulate this metric, we define $Q_\Ac = \sum_{k\in\Ac}Q_k\;\forall\;\Ac\subseteq\{1,\cdots,K\}$ and 
 use the  set functions defined in (\ref{eq:setf}) to further define $K$ functions, each over $\Iulk$, as 
\begin{eqnarray}\label{eq:setg}
g^{(\ell)}_\Gulc = \min_{\Ac\subseteq\{1,\cdots,\ell\}}\left\{Q_{\{1,\cdots,\ell\}\setminus\Ac} + f^{(\Ac)}_\Gulc \right\},\;\forall\;\Gulc\in\Iulk, 
\end{eqnarray}
where $\ell=1,\cdots,K$. Note here that for any  choice of beams and bit resolutions  $\Gulc\in\Iulk$, $g^{(\ell)}_\Gulc,\;\forall\;\ell$ is the maximal sum rate that can be achieved for users in $\Uc_\ell\define\{1,\cdots,\ell\}$,  in the presence of queue constraints (when the messages  of 
 other users $\Uc\setminus\Uc_\ell$ are known and expurgated).  Further, the rate assignment corresponding    to the desired  corner point assigns rate $R_1 =  g^{(1)}_\Gulc$ to user $1$ having the highest weight, rate 
   $R_2 =  g^{(2)}_\Gulc - g^{(1)}_\Gulc$ to user $2$ having the second highest weight and so on, till
  rate 
   $R_K =  g^{(K)}_\Gulc - g^{(K-1)}_\Gulc$ to user $K$ having the smallest weight.   We also note that 
    each $g^{(\ell)}_\Gulc$ can be efficiently computed (without brute-force search over subsets of $\Uc_\ell$) using efficient submodular function minimization routines\cite{fujiPoly}.
Next, letting $w_{K+1}=0$, we define a normalized non-negative  function  over $\Iulk$, $h:\Iulk\to\Reals_+$,   as
\begin{eqnarray}\label{eq:hfun}
h(\Gulc) = \sum_{\ell=1}^K  (w_\ell-w_{\ell+1})g^{(\ell)}_\Gulc,\;\;  \;\forall\;\Gulc\in\Iulk.
\end{eqnarray}
For any choice $\Gulc\subseteq\Omegaul: \Gulc\in\Iulk$  the beams specified by its constituent tuples are all mutually orthogonal and $h(\Gulc)$ yields the desired optimal weighted sum rate metric. 
In order to specify other constraints that any   choice of $\Gulc$ must satisfy, we associate a  cost $\epsilon_\wb+\epsilon'_{b,b_{\rm ref}} + \theta 2^b$ with each   tuple 
 $(\wb,b)\in\Omegaul$. Note here that $\theta 2^b$ denotes the energy consumed on using $b$ bit resolution ADC   whereas $\epsilon_\wb$ can account for additional circuit energy incurred on activating the RF chain and we allow for dependence of this energy term on $\wb$. Moreover, the term 
 $\epsilon'_{b,b_{\rm ref}}$ can incorporate any arbitrary (look-up-table based) switching costs incurred on changing the resolution from a given reference setting 
 $b_{\rm ref}$ to $b$ (cf. \cite{choi17}). 
Then, we define a normalized non-negative set function $c:2^\Omegaul\to\Reals_+$ such that for  any subset $\Gulc\subseteq\Omegaul$,  $c(\Gulc)$ yields the sum of costs of all tuples in $\Gulc$. Clearly $c(.)$ is a modular set function.
Thereby, we can pose our problem of interest as
\begin{eqnarray*}
\nonumber \max_{\Gulc\in\Iulk}\left\{h(\Gulc)\right\}\;\;\;
{\rm s.t.}\;\;c(\Gulc)\leq \hat{E},\;|\Gulc|\leq M' \;\;\;\;\;\;\;\;{\rm (P1)}
\end{eqnarray*}
Notice that in (P1) $\hat{E}$ denotes  given energy budget, and via the cardinality constraint on $|\Gulc|$ we have imposed  another practical constraint that only $M'$ RF chains, where $M':1\leq M'\leq M$ is a given input,  can be activated at the BS. 

In order to obtain an approximation algorithm we will first reformulate (P1). The reformulated problem is equivalent to (P1) in the sense that each feasible solution of (P1) is also feasible for the new problem, whereas each solution feasible for the latter can be mapped to one feasible for (P1) and yielding identical WSR objective. Towards this end, we  extend definition of  $h(.)$ to all subsets of $\Omegaul$, i.e., even those not in $\Iulk$.  
For any choice  $\Gulc\notin\Iulk$,  we can simply define $h(\Gulc)$  as before but doing so ignores the noise coloring caused by non-orthogonal analog beams and thus is not a physically meaningful metric although it is mathematically well defined.  
To circumvent this problem, we introduce a simple but key mathematical trick which permits us to obtain a  formulation equivalent to (P1) but {\em in which the matroid constraint  is essentially  absorbed into the objective.} In particular, for any $\Gulc\subseteq\Omegaul:\Gulc\notin\Iulk$, let us define the matrices 
  $\Lb_{\Gulc,n} = \Tb_\Gulc^{1/2}\Wb_\Gulc\Gb_n\Db_n^{1/2},\;\forall\;n=1,\cdots,N$  but where the  
 matrix  {\em $\Wb_\Gulc$ now contains as its rows only the distinct beams specified by $\Gulc$, and the matrix $\Tb_\Gulc$ is formed by using only the highest bit resolution specified in $\Gulc$ for each of its distinct beams}. With this understanding let us follow all other steps made to obtain the functions set function $h(.)$ as before. In particular, we define one set function $f^{(.)}_\Gulc$ in (\ref{eq:setf}) for each $\Gulc\subseteq\Omegaul$. Further, we define  
  $K$ set functions $g^{(\ell)}_{(.)},\ell=1,\cdots,K$ as in (\ref{eq:setg}), with each function now defined over all subsets of $\Omegaul$.  
  Let $h':2^\Omegaul\to\Reals_+$ denote the resulting extension of $h(.)$ following (\ref{eq:hfun}), which we remind is now a set function defined over all subsets of $\Omegaul$. Then, consider 
\begin{eqnarray*}
\nonumber \max_{\Gulc\subseteq\Omegaul}\left\{h'(\Gulc)\right\}\;\;\;\;\; 
{\rm s.t.}\;\;c(\Gulc)\leq \hat{E},\;\&\;|\Gulc|\leq M' \;\;\;\;\;\;\;\;\;\;\;\;{\rm (P2)}
\end{eqnarray*}
Note here that for given system dimensions ($K, N_r, |\Bc|$) each input instance of (P2) comprises of budgets $\hat{E},M'$, sets $\Wk,\Bc$, cost of each tuple $(\wb,b)\in\Omegaul$ along with all user channel matrices, transmit powers and a look-up table specifying quantization scalars as a function of  ADC bit resolutions, which together enable evaluation of the WSR objective for any choice $\Gulc$. 
We offer our key result which is proved in the appendix. 
\begin{prop}\label{prop4b}
The problem (P2) is equivalent to (P1) and itself is the maximization of a normalized monotone non-decreasing submodular set function subject to one knapsack and one cardinality   constraint.  
\end{prop}
\begin{remark}
We note that upon considering the flat fading case ($L=1$) with infinite queue sizes for all users and setting all their respective weights to be identical, our weighted sum rate metric reduces to the narrowband sum rate considered   in \cite{choi17, choiAS18}. The latter metric was optimized in \cite{choiAS18} over receive antenna subsets after assuming any arbitrarily specified but fixed bit resolution for all ADCs.  This simplified receive antenna subset selection problem itself can be shown to be NP-hard 
 which suffices to deduce that the problem in (P2) (and (P1)) is NP-hard. Thus, there is no hope of designing a polynomial-time optimal algorithm for (P2) or (P1). Our submodularity result in Proposition \ref{prop4b} assures us that the natural greedy algorithm proposed in \cite{choiAS18} for receive antenna subset selection upon explicitly modeling quantization, achieves $1-1/e$ approximation guarantee for the subset selection problem, since it is being applied on a normalized non-decreasing submodular objective subject to a cardinality constraint (cf. \cite{nemhauser78}). We note that submodularity for this latter problem without quantization (i.e.,  infinite bit resolution ADCs) has been previously established in \cite{vazeAS}.    
\end{remark}
We also remark that the joint optimization over beams and bits being considered in (P1) or (P2) requires a more sophisticated algorithm compared to the natural greedy one. In this context, note that 
  (P2) can be approximately solved with a lower complexity and a better approximation factor compared to (P1) using known algorithms for submodular maximization subject to multiple modular (knapsack) constraints. Indeed, upon applying one such multiplicative updates based algorithm     \cite{gamzu2012}  on (P2), we can deduce the following corollary. 
\begin{coro}\label{propag}
There exists a polynomial time approximation algorithm that yields a constant factor $\frac{1}{2(1+2e)}$ guarantee for (P2), i.e., for each input instance it yields a WSR that is at-least $\frac{1}{2(1+2e)}$ times the optimal WSR. 
\end{coro}


\subsection{An Enhanced Algorithm}
 We observed that there is significant scope  for improving the performance obtained by a direct application of the algorithm from \cite{gamzu2012} on (P2). 
To design our enhanced algorithm we define two set functions, one for each   constraint in (P2).
In particular, let $c':2^\Omegaul\to\Reals_+$ denote a set function such that for  any subset $\Gulc\subseteq\Omegaul$,  $c'(\Gulc)$ yields a {\em net} cost of all tuples in $\Gulc$. This net cost is determined   as the sum of normalized costs of  all distinct beams that are each present in at-least one tuple of $\Gulc$. The normalized cost associated with each such distinct beam in turn is  set to be the maximal normalized cost  (cost divided by $\hat{E}$) among all tuples of $\Gulc$ containing that beam. 
It can be verified that 
$c'(.)$ is a non-decreasing sub-modular set function over $\Omegaul$. Similarly, we define $d':2^\Omegaul\to\Reals_+$ to be a set function such that for  any subset 
$\Gulc\subseteq\Omegaul$,  $d'(\Gulc)$ equals the ratio of the number of distinct beams present across tuples of  $\Gulc$ and $M'$. Clearly, $d'(.)$  is also a non-decreasing sub-modular set function. Then, we can formulate a problem  as 
\begin{eqnarray*}
\nonumber \max_{\Gulc\subseteq\Omegaul}\left\{h'(\Gulc)\right\}\;\;\;\;\;\;\;\;\;\;\;\;\;\;\;\;\;\;\;\;\;\;\;\;\;\;\;\;\\
{\rm s.t.}\;\;c'(\Gulc)\leq 1,\;\&\;d'(\Gulc)\leq 1 \;\;\;\;\;\;\;\;\;\;\;\;\;\;\;\;\;\;{\rm (P2b)}
\end{eqnarray*}
Using arguments similar to those used to prove equivalence of (P1) and (P2), we can show that (P2) and (P2b) are equivalent.    
While replacing modular constraints by submodular ones may seem counter-intuitive, the key insight is to see equivalence of (P2) and (P2b) and noting that
\begin{eqnarray*}
c'(\Gulc)\leq c(\Gulc)/\hat{E}\;\&\;d'(\Gulc)\leq |\Gulc|/M'\;\forall\;\Gulc\subseteq\Omegaul.
\end{eqnarray*}
Thus, compared to modular constraints, using submodular constraints in (P2b) preserves equivalence while expanding the space of feasible subsets. This allows sub-optimal methods to have a better chance of escaping from poor choices. 
Next, without loss of generality, we suppose that each tuple  of $\Omegaul$ is feasible, i.e., $c'(\wb,b)\leq 1,\;\forall\;(\wb,b)\in\Omegaul$ (else we can simply remove such tuples). Further, we can also suppose that $c'(\Omegaul)>1$ and $d'(\Omegaul)>1$. Indeed, otherwise we can drop the constraints which are vacuous (i.e., met by the ground set) and in case both  $c'(\Omegaul)\leq 1$ and $d'(\Omegaul)\leq 1$ hold, an optimal solution to 
 (P2b) is trivially to choose all  beams in $\Wk$, each with the highest possible resolution in $\Bc$.

Algorithm I details the main steps of our enhanced method, where we have used $\phi$ to denote the empty set, $h'_\Gulc(\wb,b)$ to denote marginal gain $h'(\Gulc\cup(\wb,b)) - h'(\Gulc)$  (similarly $c'_\Gulc(\wb,b)$ and $d'_\Gulc(\wb,b)$). Our algorithm modifies and applies the   multiplicative updates based method, originally designed in \cite{gamzu2012} for submodular maximization subject to modular constraints, on (P2b) containing submodular constraints instead\footnote{The input parameter $\theta$ is a tuning factor which we fixed to be 2 in our simulations.}
\begin{algorithm}
\caption{Joint  Optimization}\label{alg:bit-allocJ}
\begin{algorithmic}[1]
\STATE{Set  $\Gulc=\phi$,   $V=0$ }
\STATE {Initialize $\theta\in\Reals_+, \zeta_1=1,\zeta_2=1$.} 
 \WHILE{$\zeta_1 \leq \theta\;\&\; \zeta_2 \leq \theta$ }
 \STATE{ Solve via {\em Lazy Evaluations}\\
\begin{eqnarray}\label{eq:searchstep}
\nonumber \max_{(\wb,b)\in\Omegaul\setminus\Gulc: h'_\Gulc(\wb,b)>0}\left\{\frac{h'_\Gulc(\wb,b)}{\zeta_1c'_\Gulc(\wb,b)+\zeta_2d'_\Gulc(\wb,b)}\right\}\\
{\rm s.t.}\;\; c'(\Gulc\cup(\wb,b)) \leq 1\;\&\;d'(\Gulc\cup(\wb,b)) \leq 1
\end{eqnarray}
  and let
$ (\hat{\wb},\hat{b})$ be the corresponding optimal tuple.}
 \IF {  Optimal tuple  is non-empty}
 \STATE {Augment $\Gulc\to\Gulc \cup  (\hat{\wb},\hat{b})$ and $V \to V + h'_\Gulc(\hat{\wb},\hat{b})$}
 \STATE {Update $\zeta_1=\zeta_1\theta^{c'_\Gulc(\hat{\wb},\hat{b})}$ and $\zeta_2=\zeta_2\theta^{d'_\Gulc(\hat{\wb},\hat{b})}$ }
\ELSE
 \STATE {Break}
 \ENDIF
 \ENDWHILE
\STATE {Determine $(\hat{\wb},\hat{b})=\arg\max_{ (\wb,b)\in\Omegaul }\{ h'(\wb,b) \}$}
\IF {  $h'(\hat{\wb},\hat{b})> V$}
\STATE {Set $\Gulc = (\hat{\wb},\hat{b})$.}
\ENDIF
\STATE Return $\Gulc$. 
\end{algorithmic}
\end{algorithm}
Algorithm I has several enhancements compared to the original form in \cite{gamzu2012}. 
In particular, it has an improved termination criteria (conditions in the While-Do loop) as well as a different metric in the search step (step 4)  that is derived based on the formulation in (P2b). Further,  it has an improved post-processing (steps 12-through-15). 
Notice that in each iteration, in the search step we need to determine the locally best tuple by solving (\ref{eq:searchstep}). While this entails a linear pass over $\Omegaul$, i.e., $O(|\Omegaul|)$ complexity, 
 we exploit {\em lazy evaluations} (cf. \cite{Min:Lazy}) to avoid computing metrics of several tuples that can  ascertained to  not be the locally optimal choice based on the partial ordering of marginal gains resulting from submodularity of $h'(.)$. 
 Thus, while the overall worst-case complexity of Algorithm I scales as $O(|\Omegaul|^2)$ we observed a much faster average case runtime. 
  Building upon the methodology of \cite{gamzu2012} we can prove that Algorithm I  can yield a worst-case approximation that is at-least as large as  the factor   claimed in Corollary \ref{propag}. While, we are as yet unable to establish a strictly superior performance guarantee, 
 nevertheless, as shown in the simulations our enhanced Algorithm I yields a much superior  average-case performance. 
In the following section we provide simulation results comparing our enhanced algorithm with the state of the art ones. We gratefully acknowledge the software codes provided by the authors of \cite{choiAS18} for their algorithm which allowed us to conduct a proper comparison. 
  \section{Simulation Results}\label{sec:simres}
In all the following simulations we consider a full-buffer (infinite queue sizes) scenario and assume that each user has one (omni) transmit antenna. Further, we consider the flat-fading case with $L=1$ and assume ideal channel estimation at the BS.
We compare the performance of our enhanced algorithm (Algorithm I) over practical system configurations against conventional receive antenna selection scheme (referred to here as FAS) that ignores effect of quantization, as well as the state-of-the-art quantization aware receive antenna selection scheme \cite{choiAS18} (referred to as QAFAS). The latter scheme explicitly models quantization noise but only considers antenna subset selection. In particular, it connects each selected receive antenna to a distinct RF chain and  uses a common pre-defined reference bit resolution across all ADCs.

  We begin by considering a Rayleigh fading uplink comprising of 10 users and a single BS with 128 receive antenna elements. From the available antennas  a subset of size at-most 40 can be selected and connected to the available 40 RF chains. The carrier frequency is set to  2.4 GHz, the transmission bandwidth is chosen to be 10 MHz and each user's transmit power is set to $5\;dBm$. The remaining simulation parameters such as minimum and maximum user distances in each drop, path loss exponents etc. are all as-per \cite{choiAS18}. The modeling of energy consumed by each 
 active RF chain is as-per \cite{choi17}. In Fig. \ref{fig:Rayleigh} we consider several different reference bit resolutions and plot the sum rates (or more precisely sum spectral efficiencies) of the conventional and quantization aware receive antenna selection schemes, FAS and QAFAS, respectively, along with that of a random subset selection scheme, with each scheme's performance being  averaged over several drops. We note that both FAS and QAFAS will choose 40 antennas (since there are no energy budget constraints on these two schemes) and employ the reference bit resolution across all ADCs. We then plot the averaged sum rate achieved by our enhanced algorithm which jointly optimizes the ADC bit resolution and receive antenna subset. The latter joint optimization is however subject to a sum energy constraint, where the energy budget is determined as the energy consumed by FAS and QAFAS schemes (i.e., energy expended by them to operate 40 RF chains with the reference bit resolution). In addition, we impose that the joint optimization scheme cannot employ more RF chains than the other  schemes. Finally, for each considered reference bit resolution $b$, we also impose that the dynamic range considered for  adaptive resolution spans $\max\{1,b-3\}$ through $\min\{12,b+3\}$. 
From the plot we see that significant sum-rate improvement can be achieved by our joint optimization at low to modest reference bit resolutions (for instance over $60\%$ gains at reference bit resolution 3 bits.). Interestingly,
 at larger resolutions (say 9 and above) while there is little improvement in terms of sum capacity, we have seen that our joint optimization scheme
 provides  good reduction in terms of energy consumed (even up-to $40\%$ reduction). To highlight this observation, in Table \ref{tab:compEnerratio} we tabulate the ratio of   energy consumed by the joint scheme and QAFAS, for different reference bit resolutions. Also tabulated in Table \ref{tab:compEnerratio} is the complexity ratio of joint optimization over the QAFAS scheme, where we have used the number of   sum rate evaluations as a proxy for complexity. We note here that QAFAS uses clever tricks to reduce  the burden of computing (incremental) sum rates and the impact of these are complementary to our proxy metric. We emphasize that our complexity reduction is a consequence of deducing and then exploiting submodularity in the sum rate, and the computation reduction tricks developed in \cite{choiAS18} can be used to a large extent with our scheme as well.

We now consider an mmWave uplink with carrier frequency 28 GHz, 384 receive antennas and 64 RF chains. We consider a DFT analog beamforming codebook at the BS. We remark that we have extended the quantization-aware receive antenna subset selection of \cite{choiAS18} to quantization-aware codebook subset selection,  whenever needed to generate the following curves.  We compare the performance of QAFAS and FAS with our joint optimization scheme 
 for different choices of number of users, their respective transmit powers and reference bit resolutions. 
 For each considered reference bit resolution $b$, we  impose that the dynamic range considered for  adaptive resolution in our scheme spans $\max\{1,b-4\}$ through $\min\{12,b+4\}$. This is easily done by accordingly defining the ground set $\Omegaul$ in (P2b). 

In Figs. \ref{fig:jqfb2K8} and \ref{fig:jqfb2K16} we plot the sum rate versus different user transmit powers, where for each each considered transmit power value all users transmit with that power value.  
In these figures the reference bit resolution is chosen to be 2 bits. Our joint scheme jointly optimizes the bit resolutions and codebook subset while not exceeding the energy consumed by the other two schemes and using only the available RF chains.     
In Tables \ref{tab:avgactrfb2} and \ref{tab:avgbitresb2}, we   provide the average number of active RF chains as well as the average bit resolution per active chain under our scheme. Notice here that the other two schemes will activate all 64 RF chains and use the reference bit resolution for all 64 ADCs. Moreover, in Table \ref{tab:compratiob2} we tabulate the complexity ratio of joint optimization over the QAFAS scheme, where we have again used the number of   sum rate evaluations as a proxy for complexity. 
From the plots as well as the tabulated data it is seen that joint optimization scheme has significant advantages over the state-of-art schemes and the throughput gains can be even over $40\%$.  Moreover, these gains can be achieved with a substantially reduced complexity, while consuming no greater energy than the reference schemes. 
 
In Figs. \ref{fig:jqfb4K8} and \ref{fig:jqfb4K16} we repeat the above exercise but now the reference bit resolution is chosen to be 4 bits. We see that the gains of joint optimization while somewhat reduced compared to the 2 bit reference resolution case, are still good. 
Figs. \ref{fig:jqfb8K8} and \ref{fig:jqfb8K16} on the other hand assume reference bit resolution to be 8 bits. Here there is practically no sum rate gain compared to the baseline schemes, which is because all schemes  are quite close in sum rate performance to the optimal (infinite resolution) one. Interestingly our algorithm results in significant energy savings in this regime. In Tables \ref{tab:avgactrfb8} and \ref{tab:avgbitresb8}, we   provide the average number of active RF chains as well as the average bit resolution per active chain under our scheme for these cases. In Table \ref{tab:enerratiob8} we list the energy consumption ratio of the joint optimization scheme over the QAFAS scheme.
As seen from the table there is a significant reduction in energy consumption (even exceeding $50\%$ reduction) under the joint optimization scheme, while maintaining near-optimal sum-rate  performance and with comparable complexity.  
%


\section{Conclusions and Future Work}
We proposed a  novel framework for designing algorithms to optimize  bit resolutions of analog-to-digital converters (ADCs)  as well as the choice of analog beamformers.  We demonstrated the superior performance of one algorithm we designed using the proposed framework. Several interesting avenues for future work are open. These include incorporating user scheduling wherein transmit powers (power profiles) for scheduled users are also optimized subject to additional constraints.  
\appendix
\begin{definition}
Let $\Omega$ be a ground set and $h:2^{\Omega}\to\Reals_+$  be a non-negative set function defined on the subsets of $\Omega$, that is also
normalized ($h(\emptyset)=0$) and     non-decreasing ($h(\Ac)\leq h(\Bc),\;\forall\;\Ac\subseteq\Bc\subseteq\Omega$). Then, the set function
$h(.)$ is   a {\em submodular} set function if   it satisfies,
\begin{eqnarray*}
h(\Bc\cup a)-h(\Bc)\leq h(\Ac\cup a)-h(\Ac),\\
\forall\Ac\subseteq\Bc\subseteq\Omega\;\& \;a\in\Omega\setminus\Bc.
\end{eqnarray*}
\end{definition}
\begin{definition}
  $(\Omega,\Iul)$, where $\Iul$ is collection of some subsets of $\Omega$, is said to be a {\em matroid} if 
  \begin{itemize}
 \item $\Iul$ is downward closed, i.e., $\Ac\in\Iul\;\&\;\Bc\subseteq\Ac\Rightarrow\Bc\in\Iul$
  \item For any two members $\Fc_1\in\Iul$ and $\Fc_2\in\Iul$ such that $|\Fc_1|<|\Fc_2|$, there exists
   $e\in\Fc_2\setminus\Fc_1$ such that $\Fc_1\cup\{e\}\in\Iul$. This property is referred to as the exchange property.
 \end{itemize}
\end{definition} 

\subsection{Proof of Proposition \ref{prop4b}}
We first note that any subset $\Gulc\subseteq\Omegaul$ that is feasible for (P1) is also feasible for (P2) and will satisfy $h(\Gulc)=h'(\Gulc)$.
On the other hand considering any subset $\Gulc\subseteq\Omegaul$ that is feasible for (P2) we can prune it to obtain $\tilde{\Gulc}\subseteq\Gulc$, by retaining only tuples with distinct beams and maximal bit resolutions for those beams. It is readily seen due to construction of $h'(.)$ that $h'(\tilde{\Gulc})=h'(\Gulc)$. 
 Moreover $\tilde{\Gulc}$ is feasible for (P1) with  $h'(\tilde{\Gulc})=h(\tilde{\Gulc})$. This proves the equivalence of (P1) and (P2). 
To prove the submodularity of $h'(.)$,   we first offer a  proof for  
  the case in which all queue constraints are vacuous, i.e., $Q_k=\infty\;\forall\;k\in\Uc$. We then consider   the general case with finite queues. 
In the case of infinite queues, defining $\Uc_\ell=\{1,\cdots,\ell\}\;\forall\;\ell=1,\cdots,K$, we  have 
 that 
\begin{eqnarray}\label{eq:setfA}
 f^{(\Uc_\ell)}_\Gulc = \sum_{n=1}^N\log\left|\Ib + \Lb_{\Gulc,n}^{(\Uc_\ell)}(\Lb_{\Gulc,n}^{(\Uc_\ell)})^\dag\right|,
\end{eqnarray}
where $\Lb_{\Gulc,n}^{(\Uc_\ell)}$ is formed by retaining the first $\ell$ columns of $\Lb_{\Gulc,n} = \Tb_\Gulc^{1/2}\Wb_\Gulc\Gb_n\Db_n^{1/2}$. Note that 
 the number of rows in $\Lb_{\Gulc,n}$ (and hence  $\Lb_{\Gulc,n}^{(\Uc_\ell)}$) is at-most $|\Gulc|$ since we now retain only the distinct beams across all tuples in $\Gulc$.
Further, 
\begin{eqnarray}
g^{(\ell)}_\Gulc =  f^{(\Uc_\ell)}_\Gulc,\;\forall\;\Gulc\subseteq\Omegaul,
\end{eqnarray}
and for all $\ell=1,\cdots,K$ so that 
\begin{eqnarray}\label{eq:nhdef}
h'(\Gulc) = \sum_{\ell=1}^K  (w_\ell-w_{\ell+1})f^{(\Uc_\ell)}_\Gulc,\;\;  \;\forall\;\Gulc\subseteq\Omegaul.
\end{eqnarray}
It is easy to see that $h'(.)$ is a normalized and monotone non-decreasing over $\Omegaul$. To show that this function is also submodular, we recall the definition of submodularity and consider any $\Gulc\subseteq \Gulc'\subset\Omegaul$ and
 $\eul\define(\wb,b)\in\Omegaul\setminus \Gulc'$. Notice that for any $\ell$ and $n$ we have 
\begin{eqnarray}\label{eq:setfA}
\log\left|\Ib + \Lb_{\Gulc,n}^{(\Uc_\ell)}(\Lb_{\Gulc,n}^{(\Uc_\ell)})^\dag\right|= \log\left|\Ib + (\Lb_{\Gulc,n}^{(\Uc_\ell)})^\dag\Lb_{\Gulc,n}^{(\Uc_\ell)}\right|
\end{eqnarray}
An analogous relation holds for $\Gulc'$ as well. 
Next, we observe that 
\begin{eqnarray}\label{eq:Lgprime}
\left|\Ib + (\Lb_{\Gulc',n}^{(\Uc_\ell)})^\dag\Lb_{\Gulc',n}^{(\Uc_\ell)}\right| = \left|\Ib + (\Lb_{\Gulc,n}^{(\Uc_\ell)})^\dag\Lb_{\Gulc,n}^{(\Uc_\ell)}+\Vb_{n}\right|,
\end{eqnarray}
where $\Vb_n\succeq\Zrb$ is a positive semi-definite matrix that also depends on $\Gulc,\Gulc'$ but for notational convenience we don't explicitly indicate the latter dependence. This observation stems from the fact that each beam 
 (row) in $\Wb_{\Gulc}$ is also present in $\Wb_{\Gulc'}$ and the corresponding diagonal element in $\Tb_{\Gulc}$ is no greater than the one
 in $\Tb_{\Gulc'}$. The latter fact is because increasing the bit resolution while keeping the beam fixed increases the diagonal element. \footnote{Recall that we select the maximal resolution for each beam across all tuples containing that beam in the set of interest. Each diagonal element can be expressed as $\frac{\alpha^2}{\alpha^2+\alpha(1-\alpha)\psi}$. This term is increasing in $\alpha$ for any fixed $\psi$. Hence,  keeping the beam fixed fixes the variance $\psi$ while increasing the bit resolution increases $\alpha$ and thereby the diagonal term.}  
Now suppose that the beam present in the tuple $\eul$ is some $\wb\in\Wk$. Then, we can express the incremental gain
   as 
\begin{eqnarray}
\nonumber f^{(\Uc_\ell)}_{\Gulc\cup\eul} - f^{(\Uc_\ell)}_\Gulc =\sum_{n=1}^N\log\left|\Ib + (\Lb_{\Gulc,n}^{(\Uc_\ell)})^\dag\Lb_{\Gulc,n}^{(\Uc_\ell)}+\delta\tilde{\wb}_n^\dag\tilde{\wb}_n\right|-\\
 \sum_{n=1}^N\log\left|\Ib + (\Lb_{\Gulc,n}^{(\Uc_\ell)})^\dag\Lb_{\Gulc,n}^{(\Uc_\ell)}\right|
\end{eqnarray}
where $\delta\geq 0$ is a non-negative scalar that depends on $\Gulc,\eul$ 
 and $\tilde{\wb}_n$ is a row vector containing first $\ell$ elements of  $\wb\Gb_n\Db^{1/2}_n$. Note that 
 $\delta$ is the difference between the diagonal element of $\Tb_{\Gulc\cup\eul}$ corresponding to beam $\wb$ and the 
 diagonal element of $\Tb_{\Gulc}$ corresponding to beam $\wb$.  Indeed, 
 $\delta=0$ if beam $\wb$ is already present in some tuple of $\Gulc$ with 
 a corresponding bit resolution at-least as large as the one in $\eul$.
Using this expansion with the rank-1 determinant update lemma we get 
\begin{eqnarray}\label{eq:fgdiff}
 \nonumber f^{(\Uc_\ell)}_{\Gulc\cup\eul} - f^{(\Uc_\ell)}_\Gulc = \\
\sum_{n=1}^N\log\left(1 + \delta\tilde{\wb}_n (\Ib+(\Lb_{\Gulc,n}^{(\Uc_\ell)})^\dag\Lb_{\Gulc,n}^{(\Uc_\ell)})^{-1}\tilde{\wb}_n^\dag\right). 
\end{eqnarray}
Similarly using (\ref{eq:Lgprime}) and the arguments made above, we can deduce that 
\begin{eqnarray}\label{eq:fgpdiff}
\nonumber  f^{(\Uc_\ell)}_{\Gulc'\cup\eul} - f^{(\Uc_\ell)}_{\Gulc'} = \\
\sum_{n=1}^N\log\left(1 + \delta'\tilde{\wb}_n (\Ib+(\Lb_{\Gulc,n}^{(\Uc_\ell)})^\dag\Lb_{\Gulc,n}^{(\Uc_\ell)}+\Vb_n)^{-1}\tilde{\wb}_n^\dag\right),
\end{eqnarray}
where $0\leq \delta'\leq\delta$. 
Then comparing (\ref{eq:fgdiff}) and (\ref{eq:fgpdiff}) and noting that  $\Zrb\preceq (\Ib+(\Lb_{\Gulc,n}^{(\Uc_\ell)})^\dag\Lb_{\Gulc,n}^{(\Uc_\ell)}+\Vb_n)^{-1}\preceq (\Ib+(\Lb_{\Gulc,n}^{(\Uc_\ell)})^\dag\Lb_{\Gulc,n}^{(\Uc_\ell)})^{-1}$, we get the relation
\begin{eqnarray} \label{eq:fdiffs}
 f^{(\Uc_\ell)}_{\Gulc'\cup\eul} - f^{(\Uc_\ell)}_{\Gulc'} \leq f^{(\Uc_\ell)}_{\Gulc\cup\eul} - f^{(\Uc_\ell)}_{\Gulc} . 
\end{eqnarray}
The relation in  (\ref{eq:fdiffs}) proves that for each $\ell=1,\cdots,K$, the function $f^{(\Uc_\ell)}_{(.)}$ is a submodular set function over $\Omegaul$. Then, from 
  (\ref{eq:nhdef}) we can deduce that $h'(.)$ is a linear combination of $K$ submodular set functions with non-negative combining coefficients, which proves
 that $h'(.)$ is also a submodular set function over $\Omegaul$.    

As promised above, we now consider the general case with finite queues. We will require the following two lemmas which are stated next with brief  intuitive reasoning. Their proof sketches follow later in the sequel. 
\begin{lemma}\label{lem:lemnest}
Consider any $\ell\in\{1,\cdots,K\}$ and its corresponding user set $\Uc_\ell$ along with any two subsets $\Gulc,\Gulc':\Gulc\subseteq\Gulc'\subseteq\Omegaul$.
Suppose that $\Ac_\Gulc\subseteq\Uc_\ell$ and $\Ac_{\Gulc'}\subseteq\Uc_\ell$ are the sets of users such that 
\begin{eqnarray*}
 g^{(\ell)}_\Gulc &=&  f^{(\Uc_\ell\setminus\Ac_\Gulc)}_\Gulc + Q_{\Ac_\Gulc}\\
 g^{(\ell)}_{\Gulc'} &=&  f^{(\Uc_\ell\setminus\Ac_{\Gulc'})}_{\Gulc'} + Q_{\Ac_{\Gulc'}}
\end{eqnarray*}
Then, without loss of optimality we can assume that $\Ac_\Gulc\subseteq \Ac_{\Gulc'}$. 
\end{lemma}
Note that queue constraints of users in $\Ac_\Gulc\subseteq\Uc_\ell$ are active at a  queue constrained sum rate optimal rate allocation for users in $\Uc_\ell$ when users in $\Uc\setminus\Uc_\ell$ have been expurgated and the distinct beams in $\Gulc$ are activated along with their respective maximal bit resolutions in $\Gulc$ . Lemma \ref{lem:lemnest} states that we can only have more users with active queue constraints in $\Uc_\ell$ as we add more distinct beams or improve the bit resolutions of existing ones. This is because the latter operations expand the achievable rate region. 
The other useful lemma we will invoke later is stated below.
\begin{lemma}\label{lem:mmse}
For any two user subsets $\Ac,\Bc:\Ac\subseteq\Bc\subseteq\Uc_\ell$ and  any two subsets $\Gulc,\Gulc':\Gulc\subseteq\Gulc'\subseteq\Omegaul$, we have that 
\begin{eqnarray*}
   f^{(\Bc)}_\Gulc  - f^{(\Ac)}_\Gulc
 \leq  f^{(\Bc)}_{\Gulc'}  - f^{(\Ac)}_{\Gulc'}
\end{eqnarray*}
\end{lemma}
Note that $f^{(\Bc)}_\Gulc  - f^{(\Ac)}_\Gulc$ represents the maximal sum rate (without queue constraints) that can be achieved for users in $\Bc\setminus\Ac$ when treating 
users in $\Ac$ as noise (after expurgating users in $\Uc\setminus\Bc$) and when the distinct beams in $\Gulc$ are activated along with their respective maximal bit resolutions in $\Gulc$. Lemma \ref{lem:mmse} 
 states that this sum rate must increase as we add more distinct beams or improve the bit resolutions of existing ones.

Consider any $\Gulc\subseteq \Gulc'\subset\Omegaul$ and
 $\eul\define(\wb,b)\in\Omegaul\setminus \Gulc'$. As before we will prove that the set function $g^{(\ell)}_{(.)}$ defined over $\Omegaul$ is submodular for each $\ell=\{1,\cdots,K\}$.  Consider any $\ell$ with user set $\Uc_\ell$ and let $\Ac_\Gulc$ denote users with active queue constraints under $\Gulc$. Similarly define for $\Gulc\cup\eul,\;\Gulc'\;\&\;\Gulc'\cup\eul$. 
From Lemma \ref{lem:lemnest} it follows that 
\begin{eqnarray*}
\Ac_\Gulc \subseteq\Ac_{\Gulc\cup\eul}\subseteq\Ac_{\Gulc'\cup\eul}\;\&\;
\Ac_\Gulc \subseteq\Ac_{\Gulc'}\subseteq\Ac_{\Gulc'\cup\eul}
\end{eqnarray*}
Thus, we can meaningfully define subsets  as
\begin{eqnarray*}
\Cc\define (\Ac_{\Gulc'} \cap\Ac_{\Gulc\cup\eul})\setminus\Ac_{\Gulc}\;\&\;\Dc \define \Ac_{\Gulc\cup\eul}\setminus\Ac_{\Gulc'}\\
\Fc\define \Ac_{\Gulc'\cup\eul} \setminus(\Ac_{\Gulc'}\cup \Ac_{\Gulc\cup\eul})\;\&\;\Tc \define \Uc_\ell\setminus \Ac_{\Gulc'\cup\eul}
\end{eqnarray*}
It follows that $\Ac_{\Gulc\cup\eul} = \Cc\cup \Dc\cup\Ac_\Gulc$ and $\Ac_{\Gulc'} = \Cc\cup \Ec\cup\Ac_\Gulc$, where we have set $\Ec= \Ac_{\Gulc'}\setminus\Ac_{\Gulc\cup\eul}$.
Then,  we have the   chain of inequalities given in (\ref{eq:chain1}) which establishes the desired result. 
\begin{figure*}\begin{eqnarray}\label{eq:chain1}
\nonumber  g^{(\ell)}_{\Gulc\cup\eul} - g^{(\ell)}_{\Gulc}= 
Q_\Cc +Q_\Dc + f_{\Gulc\cup\eul}^{(\Tc\cup\Fc\cup\Ec)} -  f_{\Gulc}^{(\Tc\cup\Fc\cup\Ec\cup\Cc\cup\Dc)} 
\geq  Q_\Dc + f_{\Gulc\cup\eul}^{(\Tc\cup\Fc\cup\Ec)} -  f_{\Gulc}^{(\Tc\cup\Fc\cup\Ec\cup\Dc)}\\
\nonumber = Q_\Dc + ( f_{\Gulc\cup\eul}^{(\Tc\cup\Fc)} -  f_{\Gulc}^{(\Tc\cup\Fc)}) +  ( f_{\Gulc\cup\eul}^{(\Ec|\Tc\cup\Fc)} -  f_{\Gulc}^{(\Ec|\Tc\cup\Fc)})
 - f_{\Gulc}^{(\Dc|\Tc\cup\Fc\cup\Ec)}\\
\nonumber \geq Q_\Dc + ( f_{\Gulc\cup\eul}^{(\Tc\cup\Fc)} -  f_{\Gulc}^{(\Tc\cup\Fc)})  
 - f_{\Gulc'}^{(\Dc|\Tc\cup\Fc\cup\Ec)}\\
\nonumber\geq Q_\Dc + ( f_{\Gulc'\cup\eul}^{(\Tc\cup\Fc)} -  f_{\Gulc'}^{(\Tc\cup\Fc)}) 
 - f_{\Gulc'}^{(\Dc|\Tc\cup\Fc)} = (Q_\Dc +  f_{\Gulc'\cup\eul}^{(\Tc\cup\Fc)}) -  f_{\Gulc'}^{(\Dc\cup\Tc\cup\Fc)}\\
\geq g^{(\ell)}_{\Gulc'\cup\eul} - g^{(\ell)}_{\Gulc'}
\end{eqnarray}
\end{figure*}
In (\ref{eq:chain1})  we have used $f_{\Gulc\cup\eul}^{(\Ec|\Tc\cup\Fc)}=f_{\Gulc\cup\eul}^{(\Ec\cup\Tc\cup\Fc)}-f_{\Gulc\cup\eul}^{(\Tc\cup\Fc)}$ (similarly for other such terms).
To derive the first inequality in (\ref{eq:chain1}) we have simply used definition of $g^{(\ell)}_{\Gulc}$ to upper bound it as $g^{(\ell)}_{\Gulc}= Q_{\Ac_\Gulc}+f_{\Gulc}^{(\Tc\cup\Fc\cup\Ec\cup\Cc\cup\Dc)} \leq Q_{\Ac_\Gulc}+ Q_{\Cc} + f_{\Gulc}^{(\Tc\cup\Fc\cup\Ec\cup\Dc)}$ and to derive the second equality we have used the chain rule of mutual information. To derive the second inequality 
 we have invoked Lemma \ref{lem:mmse} to deduce that $( f_{\Gulc\cup\eul}^{(\Ec|\Tc\cup\Fc)} -  f_{\Gulc}^{(\Ec|\Tc\cup\Fc)})\geq 0$ and that
  $f_{\Gulc'}^{(\Dc|\Tc\cup\Fc\cup\Ec)} \geq f_{\Gulc}^{(\Dc|\Tc\cup\Fc\cup\Ec)}$.  
To derive the third inequality we have reused the submodularity of $f^{\Ac}_{(.)}$ for any user set $\Ac$ that we proved earlier
 along with the fact that $f_{\Gulc'}^{(\Dc|\Tc\cup\Fc\cup\Ec)}\leq f_{\Gulc'}^{(\Dc|\Tc\cup\Fc)} $. The latter fact is simply because removing more interfering users will increase achievable sum-rate of users in $\Dc$. Finally, the last inequality follows upon using the 
 fact that $g^{(\ell)}_{\Gulc'} = f_{\Gulc'}^{(\Dc\cup\Tc\cup\Fc)} + Q_{\Ac_\Gulc\cup\Cc\cup\Ec} $ and 
definition of $g^{(\ell)}_{\Gulc'\cup\eul}$ to deduce $g^{(\ell)}_{\Gulc'\cup\eul}\leq Q_{\Ac_\Gulc\cup\Cc\cup\Ec\cup\Dc} +  f_{\Gulc'\cup\eul}^{(\Tc\cup\Fc)} $.

\endproof

\subsection{Proof of Lemma \ref{lem:mmse}}
Let us first consider the case when $\Gulc$ and $\Gulc'$ have the same set of distinct beams in their respective constituent tuples. 
Then we can write
\begin{eqnarray*}
f_{\Gulc'}^{(\Bc)} - f_{\Gulc'}^{(\Ac)} = \\
\sum_{n=1}^N\log\left|\Ib + (\Lb_{\Gulc,n}^{(\Bc\setminus\Ac)})^\dag\Sigmab(\Ib + \Sigmab\Lb_{\Gulc,n}^{(\Ac)}(\Lb_{\Gulc,n}^{(\Ac)})^\dag\Sigmab)^{-1}\Sigmab\Lb_{\Gulc,n}^{(\Bc\setminus\Ac)}\right|,
\end{eqnarray*} 
where we have used the fact that $\Gulc'$ has  at-least as large a maximal resolution for each distinct beam as compared to $\Gulc$, so that  
$\Sigmab \define \Tb_{\Gulc'}^{1/2}\Tb_{\Gulc}^{-1/2}\succeq\Ib$.
Clearly, since $\Sigmab$ is diagonal, $\Sigmab^2\succeq\Ib$ so we have the following set of relations.
\begin{eqnarray*}
f_{\Gulc'}^{(\Bc)} - f_{\Gulc'}^{(\Ac)} = \\
= \sum_{n=1}^N\log\left|\Ib + (\Lb_{\Gulc,n}^{(\Bc\setminus\Ac)})^\dag(\Sigmab^{-2} + \Lb_{\Gulc,n}^{(\Ac)}(\Lb_{\Gulc,n}^{(\Ac)})^\dag)^{-1}\Lb_{\Gulc,n}^{(\Bc\setminus\Ac)}\right|\\
\geq \sum_{n=1}^N\log\left|\Ib + (\Lb_{\Gulc,n}^{(\Bc\setminus\Ac)})^\dag(\Ib + \Lb_{\Gulc,n}^{(\Ac)}(\Lb_{\Gulc,n}^{(\Ac)})^\dag)^{-1}\Lb_{\Gulc,n}^{(\Bc\setminus\Ac)}\right|\\
= f_{\Gulc}^{(\Bc)} - f_{\Gulc}^{(\Ac)} 
\end{eqnarray*} 
Then, consider the general case where $\Gulc'$ can also include additional distinct beams than $\Gulc$. From the result we obtained above we can deduce that, starting from $\Gulc'$ and decreasing the maximal resolution of any beam (equivalently decreasing the corresponding diagonal entry of $\Tb_{\Gulc'}$)  while keeping all other beams and their respective maximal resolutions unchanged, will decrease the sum rate of users in $\Bc\setminus\Ac$ (when treating users in $\Ac$ as noise). Moreover,  when that entry becomes zero the resulting sum rate is the one obtained upon removing all tuples containing that beam from $\Gulc'$.
 This demonstrates that as we morph $\Gulc'$ to $\Gulc$ the sum rate of users in $\Bc\setminus\Ac$ is non-increasing, which proves the lemma. 
\subsection{Proof of Lemma \ref{lem:lemnest}}
We will prove this result via contradiction. Let $\Sc=\Ac_{\Gulc}\setminus\Ac_{\Gulc'}$ such that $\Sc\neq\phi$, i.e., $\Sc$ is not empty  and let
 $\Sc'=\Ac_{\Gulc'}\setminus\Ac_{\Gulc}$. Further, define $\Tc=\Ac_{\Gulc}\cap\Ac_{\Gulc'}$ and 
$\Ec = \Uc_\ell\setminus(  \Tc\cup\Sc\cup\Sc')$. Thus,  
under $\Gulc$, we can parse $\Uc_\ell$ into users with active and inactive queue constraints, respectively, as $\Ac_\Gulc\cup (\Sc'\cup\Ec) $ while  under $\Gulc'$, we can similarly parse $\Uc_\ell$ as $\Ac_{\Gulc'}\cup (\Sc\cup\Ec) $.
By the definition of  $g^{(\ell)}_\Gulc$, we have that 
\begin{eqnarray*}
 g^{(\ell)}_\Gulc =  f^{(\Sc'\cup\Ec)}_\Gulc + Q_{\Tc} + Q_{\Sc} 
 \leq   f^{(\Sc'\cup\Ec)}_\Gulc + f^{(\Sc|\Sc'\cup\Ec)}_\Gulc+ Q_{\Tc}  
 \end{eqnarray*}
which yields that 
\begin{eqnarray*} 
  Q_{\Sc}\leq f^{(\Sc|\Sc'\cup\Ec)}_\Gulc
  \end{eqnarray*}
  Combining this with Lemma \ref{lem:mmse} we get that 
  \begin{eqnarray}\label{eq:sleqq0}
  Q_{\Sc} \leq  f^{(\Sc|\Sc'\cup\Ec)}_{\Gulc} \leq f^{(\Sc|\Sc'\cup\Ec)}_{\Gulc'}
  \leq  f^{(\Sc|\Ec)}_{\Gulc'}
  \end{eqnarray}
  where for the last inequality we have used the fact that removing interfering users will improve the sum rate of users in $\Sc$. 
  Then, by the definition of  $g^{(\ell)}_{\Gulc'}$, we have that 
\begin{eqnarray*}
 g^{(\ell)}_{\Gulc'} =  f^{(\Sc\cup\Ec)}_{\Gulc'} + Q_{\Tc} + Q_{\Sc'}
 \leq    f^{(\Ec)}_{\Gulc'} + Q_{\Sc}+ \Qc_{\Sc'}+ Q_{\Tc}  
 \end{eqnarray*} 
 which means that 
  \begin{eqnarray}  \label{eq:sleqq}
  Q_{\Sc}\geq f^{(\Sc|\Ec)}_{\Gulc'}
  \end{eqnarray}
Clearly if (\ref{eq:sleqq}) holds with strict inequality we have a contradiction from (\ref{eq:sleqq0}). On the other hand, if (\ref{eq:sleqq}) holds with equality we can also express
 \begin{eqnarray*}
 g^{(\ell)}_{\Gulc'} =  f^{(\Ec)}_{\Gulc'} + Q_{\Sc}+ \Qc_{\Sc'}+ Q_{\Tc}  
 \end{eqnarray*}
so that the set of users in $\Uc_\ell$ with active queue constraints under  a sum-rate optimal allocation and  $\Gulc'$
 can also be $\Sc\cup\Sc'\cup\Tc$ which subsumes $\Ac_{\Gulc}$ and hence proves the lemma.

\begin{figure}[t]
 \begin{minipage}[t]{\linewidth}
    \centering
    \includegraphics[width=.71\linewidth]{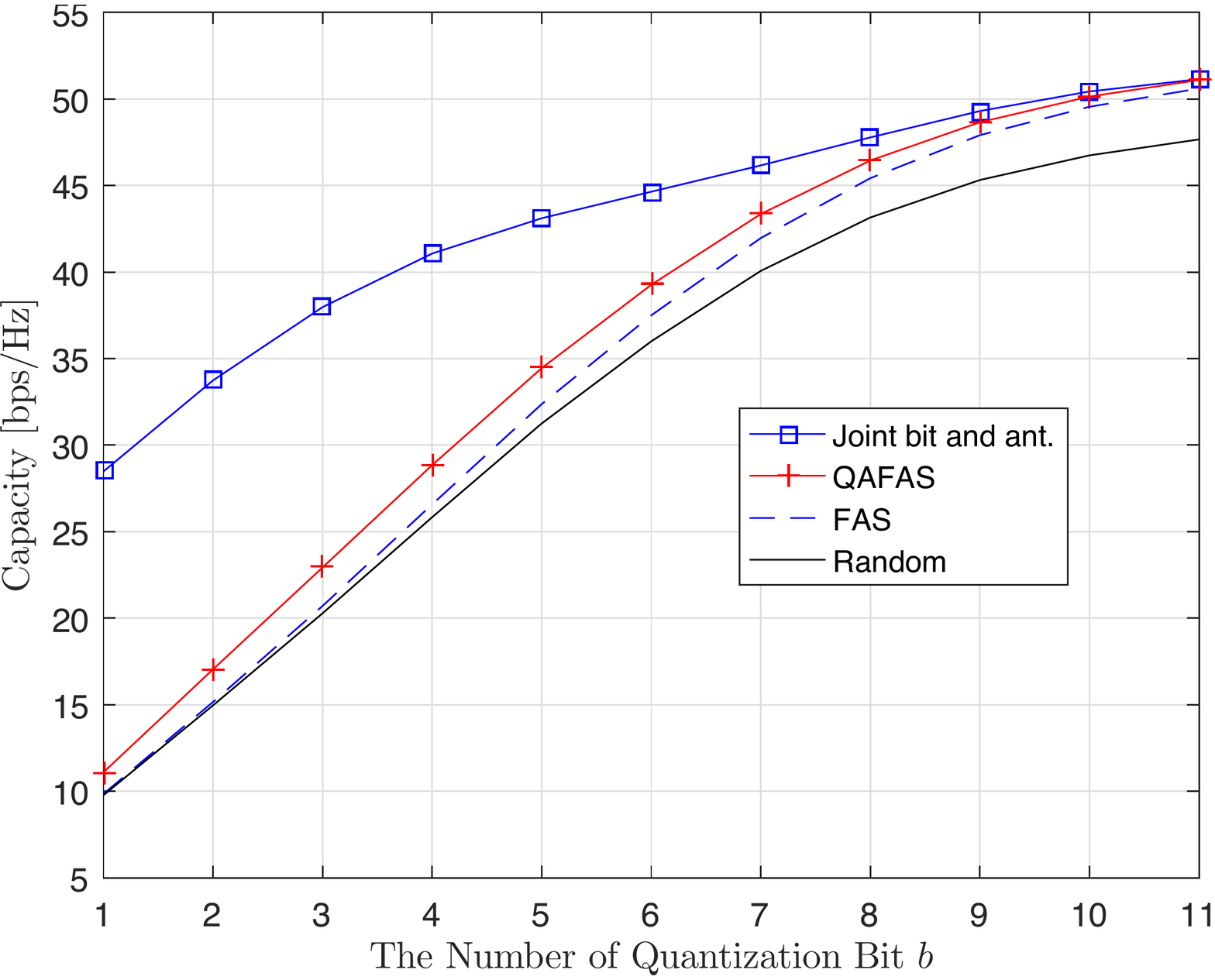}
    \caption{Sum Rate versus reference bit resolution}\label{fig:Rayleigh}\vspace{-.75cm}
\end{minipage}\vspace{-.75cm}
\end{figure}

\begin{table}[]
\centering
\caption{Energy and Complexity Ratios  }
\label{tab:compEnerratio}
\begin{tabular}{l|lllllllllll|}
& b=1 &   b=3 &   b=5 &     b=7 &   b=9 &   b=11      \\\hline
Energy Ratio & 0.98  &      0.99 &      0.98  &      0.97 &      0.83 &      0.59 \\
 Complexity Ratio &  0.67 &      0.77 &     0.94  &    1.08 &     1.18  &     1.02  \\
 \end{tabular}
\end{table}


\begin{table}[]
\centering
\caption{Avg. number of active chains for $b_{\rm ref}=2$ }
\label{tab:avgactrfb2}
\begin{tabular}{l|llllll|}
& -5 dBm & 0 dBm & 5 dBm & 10 dBm & 15 dBm & 20 dBm\\\hline
 K=8   & 51.61  & 50.52  & 49.00  & 46.86  & 43.57  & 41.48 \\
 K=16  & 51.95 &  50.36 &  48.59  & 46.57 &  43.50  & 40.39 
\end{tabular}
\end{table}

\begin{table}[]
\centering
\caption{Avg. bit resolution per active chain for $b_{\rm ref}=2$ }
\label{tab:avgbitresb2}
\begin{tabular}{l|llllll|}
   & -5 dBm & 0 dBm & 5 dBm & 10 dBm & 15 dBm & 20 dBm \\\hline
K=8   & 4.19   & 4.33   & 4.50  &  4.73   & 5.11   & 5.37 \\
 K=16   & 4.17  &  4.35 &   4.56  &  4.79  &  5.15 &  5.51 
\end{tabular}
\end{table}

\begin{table}[]
\centering
\caption{Complexity Ratios for $b_{\rm ref}=2$ }
\label{tab:compratiob2}
\begin{tabular}{l|llllll|}
& -5 dBm & 0 dBm & 5 dBm & 10 dBm & 15 dBm & 20 dBm\\\hline
 K=8   & 0.25 &   0.25   & 0.29  &  0.28  &  0.28  &  0.28  \\
 K=16   & 0.23 &   0.24 &   0.25  &  0.25  &  0.26  &  0.26  
\end{tabular}
\end{table}

\begin{table}[]
\centering
\caption{Avg. number of active chains for $b_{\rm ref}=8$ }
\label{tab:avgactrfb8}
\begin{tabular}{l|llllll|}
& -5 dBm & 0 dBm & 5 dBm & 10 dBm & 15 dBm & 20 dBm\\\hline
 K=8   & 64.00  & 64.00  & 64.00 &  64.00  & 64.00  & 63.70 \\
 K=16   & 64.00 &  64.00 &  64.00 &  64.00 &  64.00 &  63.88
\end{tabular}
\end{table}

\begin{table}[]
\centering
\caption{Avg. bit resolution per active chain for $b_{\rm ref}=8$ }
\label{tab:avgbitresb8}
\begin{tabular}{l|llllll|}
   & -5 dBm & 0 dBm & 5 dBm & 10 dBm & 15 dBm & 20 dBm \\\hline
K=8   & 5.40   & 5.58  &  5.69  &  6.02   & 6.18  &  6.53 \\
 K=16  & 5.42  &  5.56 &   5.86 &   6.05 &   6.46 &   6.84 
\end{tabular}
\end{table}

\begin{table}[]
\centering
\caption{Energy Ratios for $b_{\rm ref}=8$ }
\label{tab:enerratiob8}
\begin{tabular}{l|llllll|}
& -5 dBm & 0 dBm & 5 dBm & 10 dBm & 15 dBm & 20 dBm\\\hline
 K=8   & 0.41  &  0.45   & 0.48  &  0.58  &  0.63  &  0.75  \\
 K=16   & 0.41  &  0.43 &   0.50  &  0.53 &   0.64  &  0.77  
\end{tabular}
\end{table}

%

\begin{figure}
\centering 
\begin{minipage}[c]{0.71\linewidth}
  \centering \vspace{-.25cm} 
    \includegraphics[width=0.95\linewidth]{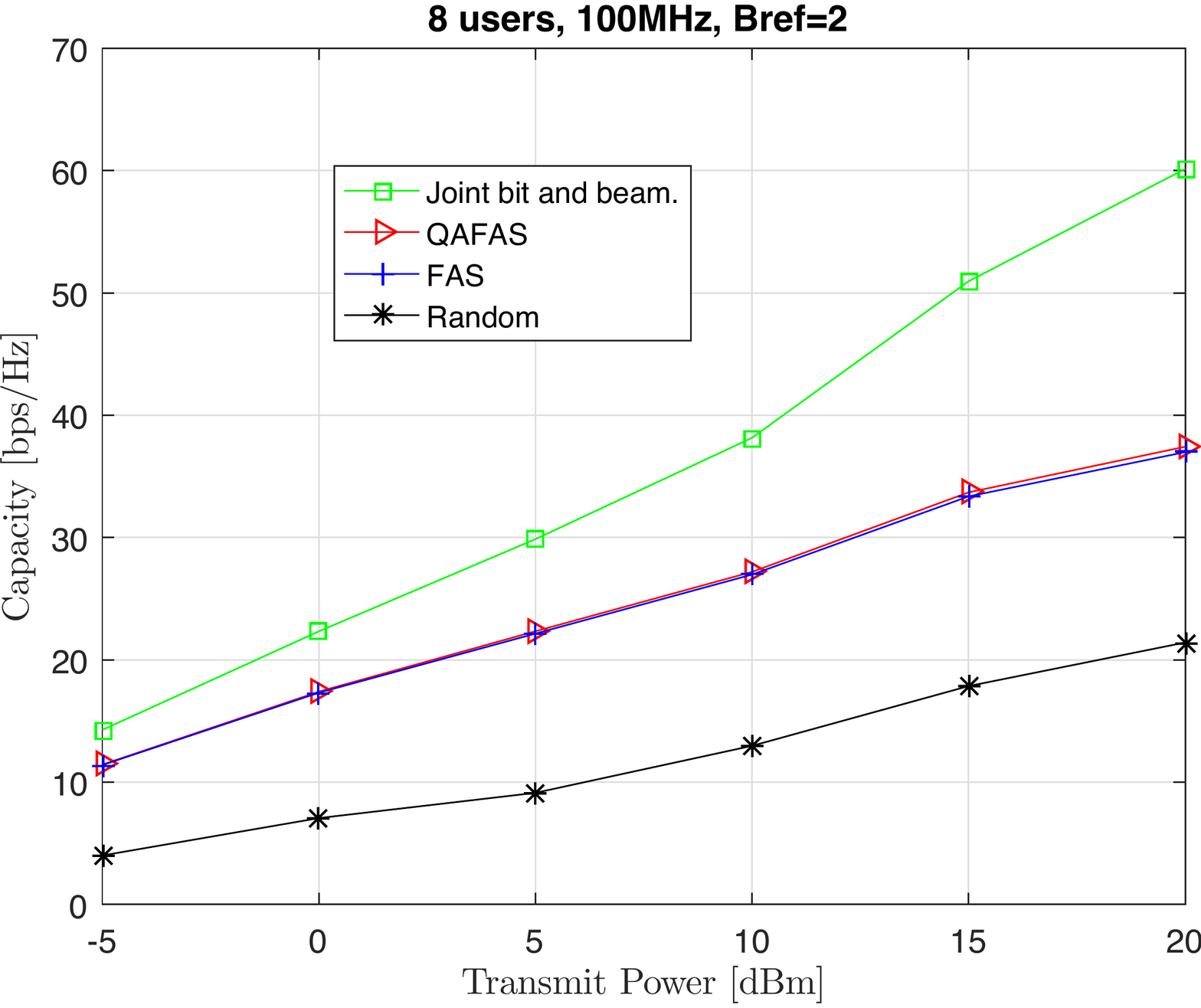}
\caption{\scriptsize Sum rate versus transmit powers for $K=8$ users and
 $\rm{b_{ref}}=2$ bits.}\label{fig:jqfb2K8}
\end{minipage}\hfill
\begin{minipage}[c]{0.71\linewidth}
  \centering  \vspace{-.75cm}
  \includegraphics[width=0.95\linewidth]{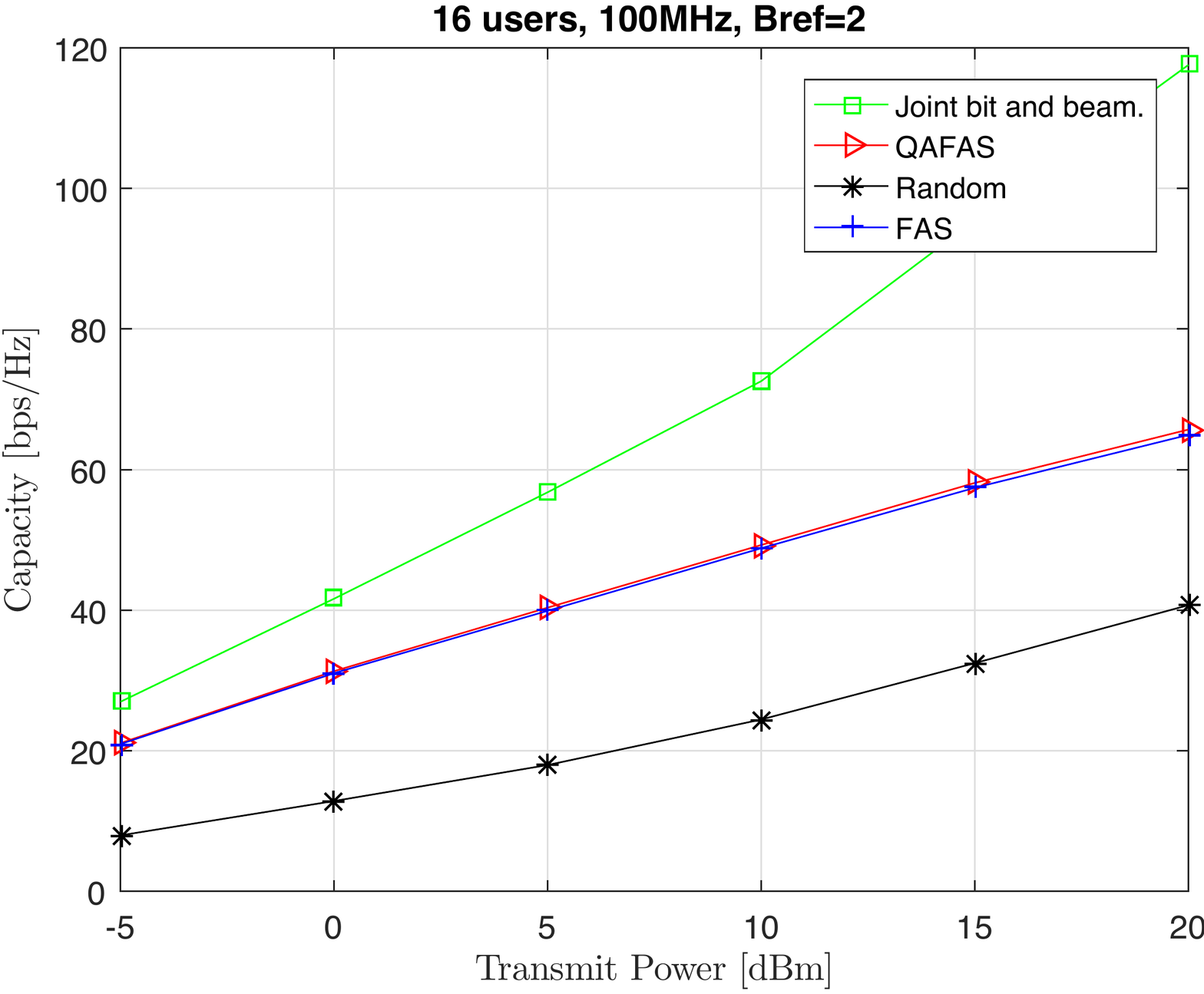}
\caption{\scriptsize Sum rate versus transmit powers $K=16$ users and
 $\rm{b_{ref}}=2$ bits.}\label{fig:jqfb2K16}
\end{minipage} 
\end{figure}


\begin{figure}
\begin{minipage}[t]{0.71\linewidth}
  \centering \vspace{-.25cm} 
    \includegraphics[width=.95\linewidth]{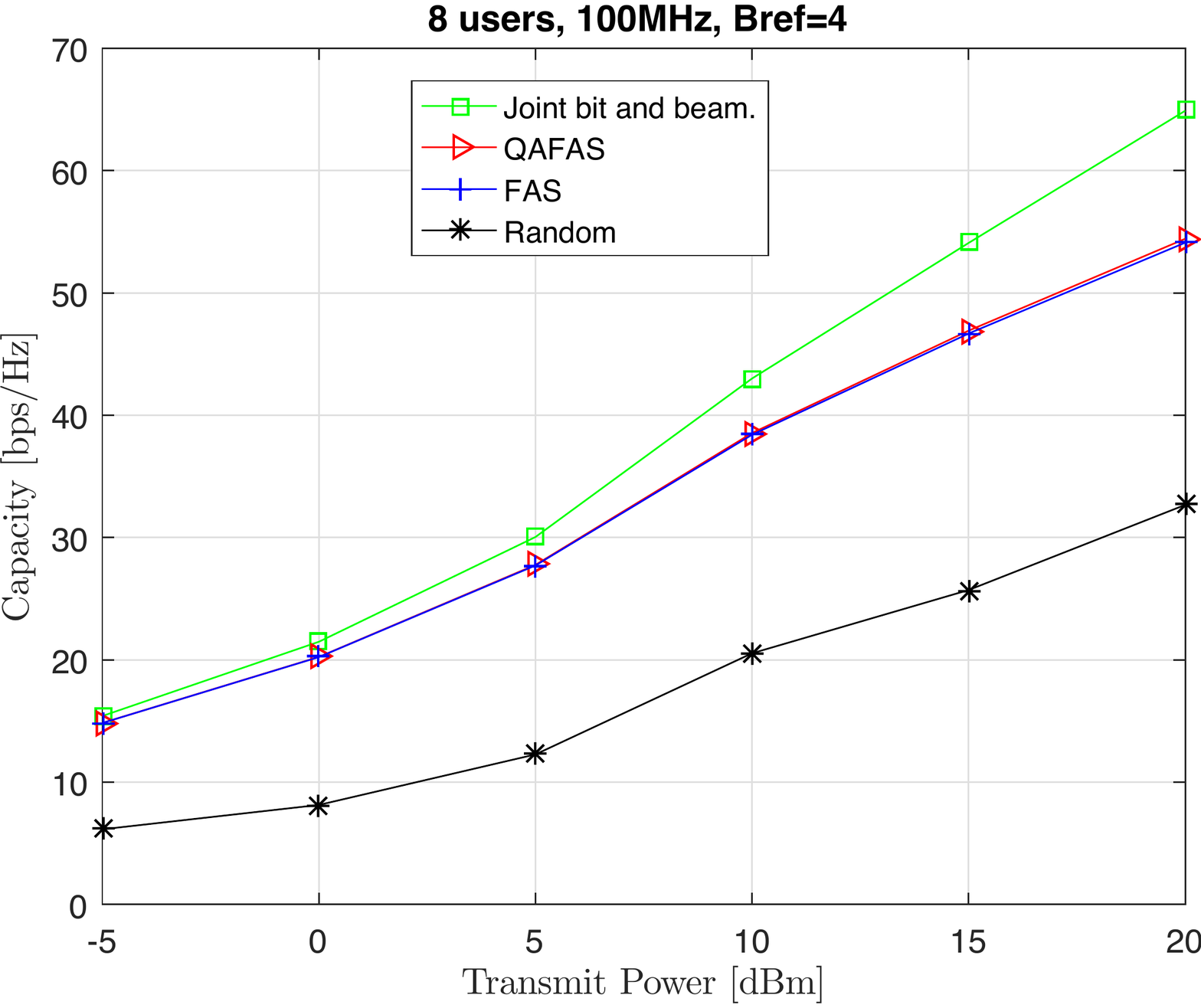}
\caption{Sum rate versus transmit powers $K=8$ users and
 $\rm{b_{ref}}=4$ bits.}\label{fig:jqfb4K8}
\end{minipage}\hfill
\begin{minipage}[t]{0.71\linewidth}
  \centering  \vspace{-.75cm}
  \includegraphics[width=.95\linewidth]{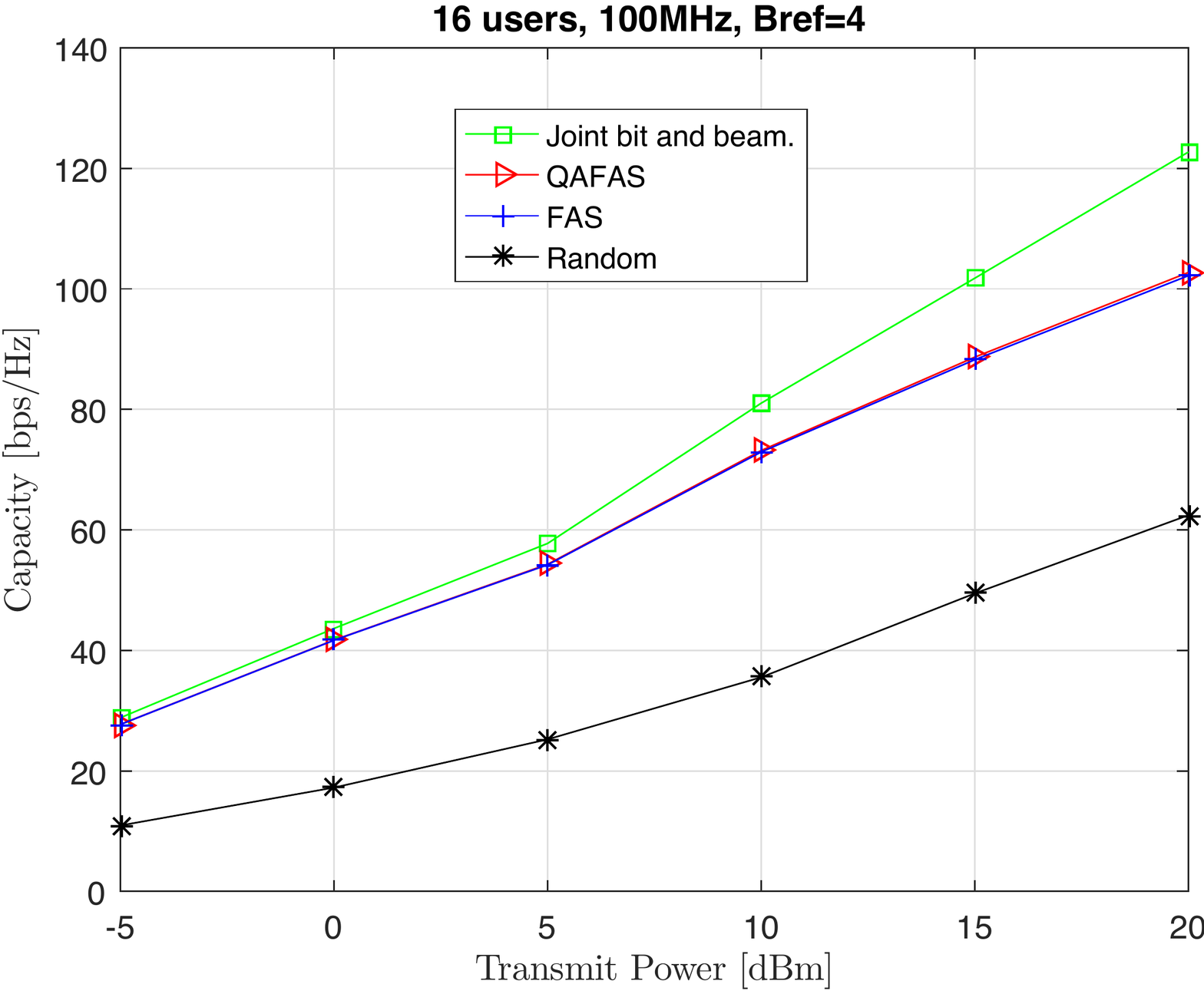}
\caption{Sum rate versus transmit powers $K=16$ users and
 $\rm{b_{ref}}=4$ bits.}\label{fig:jqfb4K16}
\end{minipage} 
\end{figure}

%

\begin{figure}
\begin{minipage}[t]{0.71\linewidth}
  \centering \vspace{-.25cm} 
    \includegraphics[width=.95\linewidth]{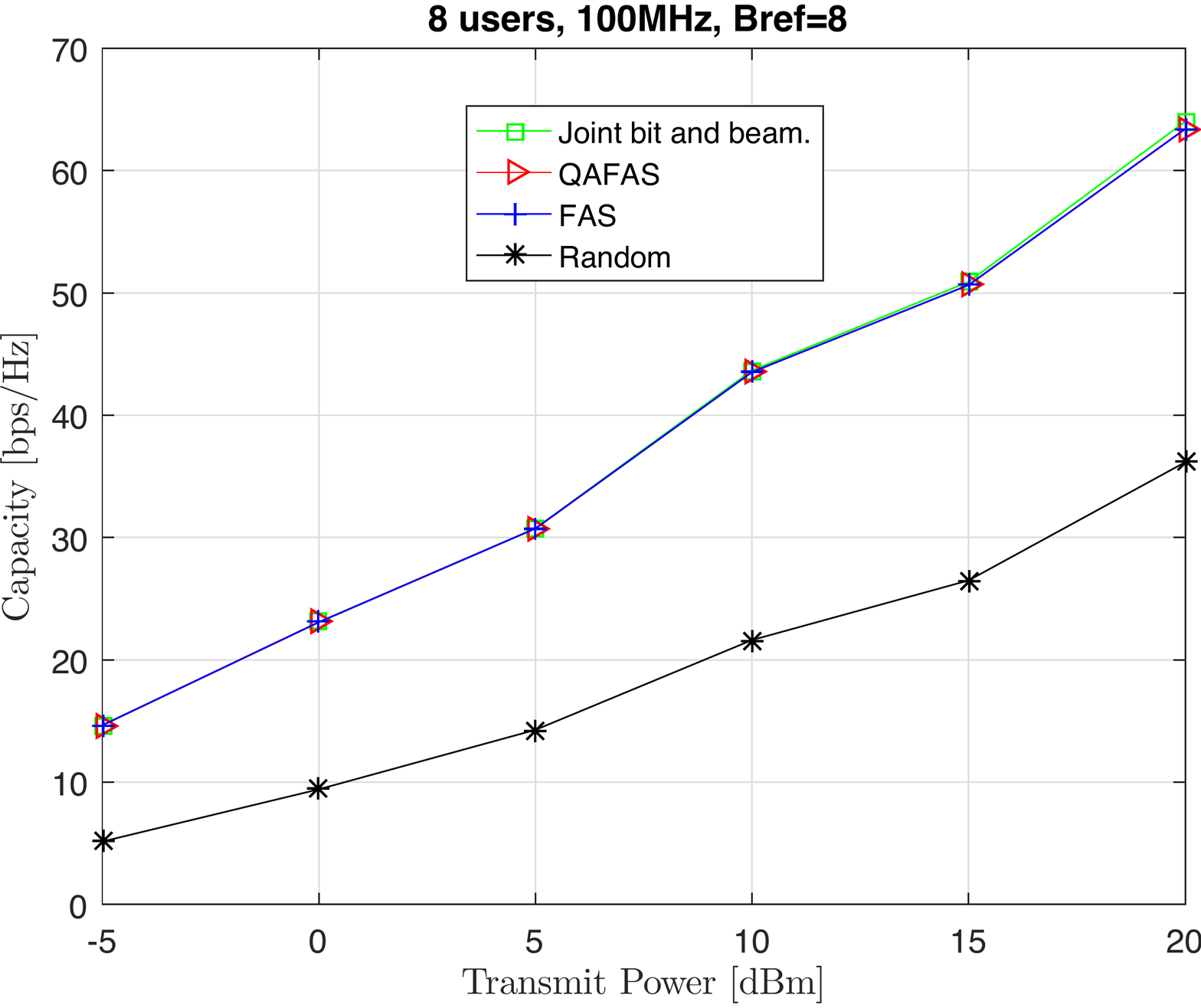}
\caption{Sum rate versus transmit powers $K=8$ users and
 $\rm{b_{ref}}=8$ bits.}\label{fig:jqfb8K8}
\end{minipage}\hfill
\begin{minipage}[t]{0.71\linewidth}
  \centering  \vspace{-.75cm}
  \includegraphics[width=.95\linewidth]{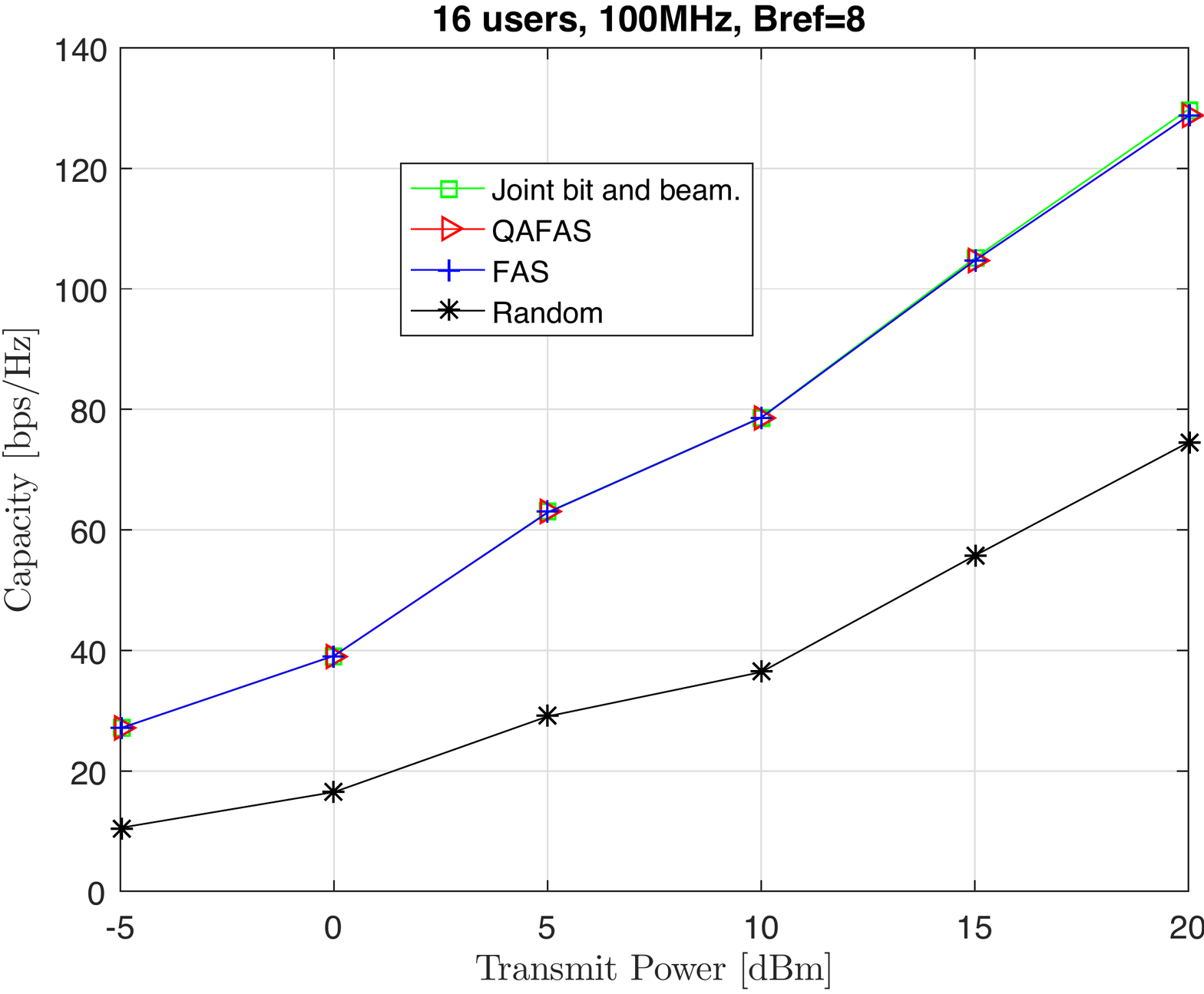}
\caption{Sum rate versus transmit powers $K=16$ users and
 $\rm{b_{ref}}=8$ bits.}\label{fig:jqfb8K16}
\end{minipage} 
\end{figure}

%

 \bibliography{Prasadv2}
\bibliographystyle{ieeetr}

\end{document}